\documentclass[twocolumn,5p]{elsarticle}
\makeatletter
\def\ps@pprintTitle{%
 \let\@oddhead\@empty
 \let\@evenhead\@empty
 \def\@oddfoot{}%
 \let\@evenfoot\@oddfoot}
\makeatother

\usepackage{multibib}
\newcites{data}{Citations Belonging with the Data}

\usepackage{calc}
\usepackage{hhline}
\usepackage{lineno,hyperref}
\usepackage{graphicx}
\usepackage{amssymb}
\usepackage{arrayjobx}
\usepackage{trimspaces}
\usepackage{hyperref}
\usepackage{forloop}
\usepackage{array,multirow,graphicx}
\usepackage{float}
\usepackage{url}
\usepackage[figuresright]{rotating}
\usepackage[section]{placeins}
\usepackage{pdfpages}

\def\trimspace#1{\trim@spaces@in{#1}}

\extrafloats{1000}
\begin{document}
\newcommand{\numberofpapercases}{six}
\newcommand{\numberofactualcases}{six}
\newcommand{\numberofactualcasecompanies}{four}
\newcommand{\numberofactualpractices}{168}

\begin{frontmatter}
\title{Source Data for the Focus Area Maturity Model for API Management}

\author[1,2]{Max Mathijssen}
\author[1,2]{Michiel Overeem}
\author[1,3]{Slinger Jansen}

\address[1]{m.mathijssen@students.uu.nl, \{m.overeem, slinger.jansen\}@uu.nl, Utrecht University, Princetonplein 5, 3584CH Utrecht, The Netherlands}
\address[2]{\{max.mathijssen, michiel.overeem\}@afas.nl, AFAS Software, Philipsstraat 9, 3833 LC Leusden, The Netherlands}
\address[3]{\ LUT University, Yliopistonkatu 34, 53850 Lappeenranta, Finland}

\begin{abstract}
API Management is the design, publication, and deployment of APIs by an organization for (external) developers to consume. API Management encompasses capabilities such as controlling API lifecycles, access and authentication to APIs, monitoring, throttling and analyzing API usage, as well as providing security and documentation. This dataset describes the API Management Focus Area Maturity Model (API-m-FAMM). In a structured manner, this model aims to support organizations that expose their API(s) to third-party developers in their API management activities. The model is developed through a Systematic Literature Review (SLR), expert interviews, and case studies. As a result, a model consisting of 80 practices and 20 capabilities, which are assigned to 6 focus areas. The practices the model comprises are described by a practice code, name, description, conditions for implementation, as well as the associated literature in which the practice was originally identified. Capabilities and focus areas are described by a code, description and, optionally, the associated literature in which it was originally identified. Using the API-m-FAMM, organizations may evaluate, improve upon and assess the degree of maturity their business processes regarding the topic of API management have.

\end{abstract}

\end{frontmatter}

\section{Data Specifications Table}

\begin{table}[htb]
\centering
\footnotesize
\label{DataSpecificationTable}
\begin{tabular}{|l|p{10cm}|}
\hline
\textbf{	Subject	}&	Management of Technology and Innovation.	\\\hline
\textbf{	Specific subject area	}&	A focus area maturity model for API management.	\\\hline
\textbf{	Type of data	}&	Text, literature references, and tables.	\\\hline
\textbf{	How data were acquired	}&	Systematic literature review and expert interviews.	\\\hline
\textbf{	Data format	}&	Raw, analyzed, and evaluated.	\\\hline
\textbf{	Parameters for data collection	}&	The collected practices had to fit strict requirements in terms of having to be executable, implementable, and easily understandable by practitioners that are involved with API management within their organization.	\\\hline
\textbf{	Description of data collection	}&	The initial data was collected through a SLR \cite{mathijssen2020identification}. Initially, the data was grouped according to topical similarity. Practices were categorized, analyzed and verified through discussion sessions with all involved researchers, inter-rater agreement and information gathered from grey literature. Capabilities and practices were then evaluated through 11 expert interviews. For information on selection of the practitioners, we refer to the related research article \textit{(to be published)}. If at least 2 or more practitioners found a practice relevant and useful, they became a part of the collection. Additionally, six discussion sessions among the researchers were conducted, during which all suggested changes (i.e. removal, addition, and relocation of practices and capabilities) were discussed, interpreted, and processed. The resulting practices and capabilities were then evaluated with 3 experts whom were previously interviewed.
Finally five case studies were conducted to evaluate different software products.
\\\hline
\textbf{	Data source location	}&	All included source literature can be reviewed in the associated research article~\cite{mathijssen2020identification}.	\\\hline
\textbf{	Related research article	}&	Mathijssen, M., Overeem, M., \& Jansen, S. (2020). Identification of Practices and Capabilities in API Management: A Systematic Literature Review. arXiv preprint arXiv:2006.10481.\\\hline
\end{tabular}
\end{table}

\onecolumn

\section{Introduction}
\label{sec:introduction}

This data set describes the API Management Focus Area Maturity Model (API-m-FAMM). 
The model supports organizations that expose their API(s) to third-party developers, in a structured manner, in their API management activities. 
Using the API-m-FAMM, organizations may evaluate, improve upon and assess the degree of maturity their business processes regarding the topic of API management have.

We define API Management as an activity that enables organizations to design, publish and deploy their  APIs  for  (external)  developers  to  consume.  API  Management encompasses capabilities  such as  controlling  API  lifecycles,  access  and  authentication  to  APIs,  monitoring,  throttling and  analyzing  API  usage,  as  well  as  providing  security  and  documentation.

\begin{itemize}
    \item The data may be used by API management researchers for evaluation, validation and extension of the model.
    \item The data can be used by focus area maturity researchers to establish the vocabulary used in the field.
    \item The data can be used by researchers as a basis for future research work in the domains of API management, versioning and evolution.
    \item The data is reusable by consultants and practitioners to assess whether they have implemented a practice fully.
\end{itemize}

The research approach is explained in Section~\ref{sec:design}.
Section~\ref{sec:apimfamm} describes the final API-m-FAMM in full detail.
The different intermediate versions are described in Sections~\ref{sec:version01}, \ref{sec:version02}, \ref{sec:version03}, \ref{sec:version04}, \ref{sec:version05}, and \ref{sec:version10}.

\section{Experimental Design, Materials, and Methods}
\label{sec:design}

\begin{figure*}[!h]
  \centering
  \includegraphics[page=1, clip, trim=1.0cm 12.5cm 2.1cm 0.8cm, width=\textwidth]{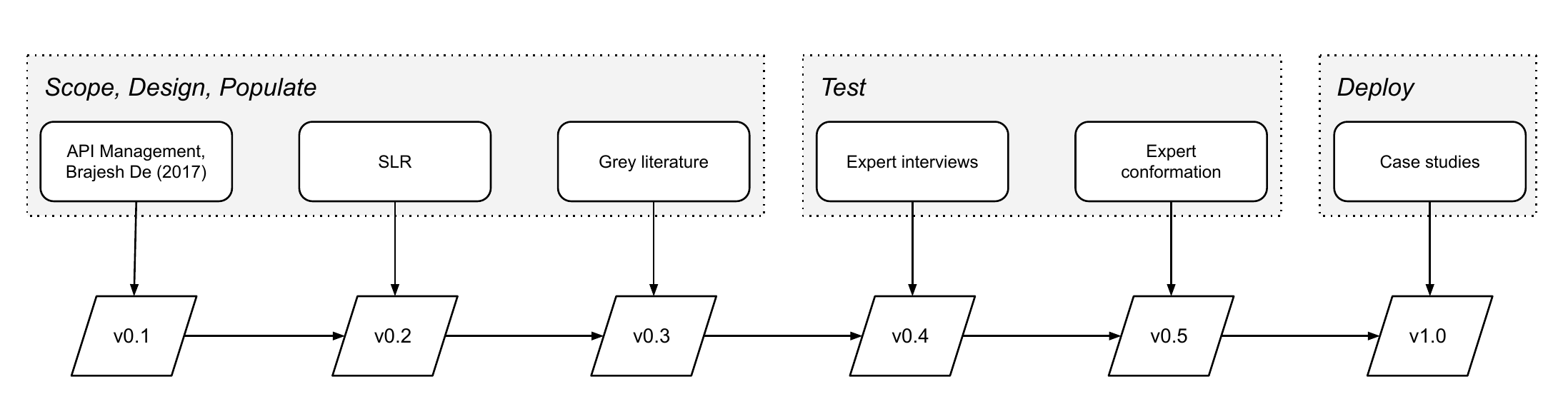}
  \caption{The steps that were executed in constructing the API-m-FAMM and its various intermediate versions.}
  \label{fig:research-steps}
\end{figure*}

The Focus Area Maturity Model is constructed using the design methodology of \cite{van2010design} and \cite{de2005understanding}.
The development of the FAMM is done in five phases: \emph{Scope}, \emph{Design}, \emph{Populate}, \emph{Test}, and \emph{Deploy}.
These phases are executed through a SLR, expert interviews, case studies, and numerous discussions among the authors.
Between the execution of every method, the authors discussed the state of the model until consensus was reached on its contents and structure.
This was done using online \textit{Card Sorting}~\citep{nielsen1995}, with \textit{Google Drawings} as a tool. 
Figure~\ref{fig:research-steps} shows which methods were used in each phase, by linking them to the different intermediate versions of the API-m-FAMM.
The intermediate versions including a changelog are described in Sections~\ref{sec:version01}, \ref{sec:version02}, \ref{sec:version03}, \ref{sec:version04}, \ref{sec:version05}, and \ref{sec:version10}.

\subsection{Scope, Design, Populate Phases}

The initial data was acquired through the SLR as described in \cite{mathijssen2020identification}.
Based on this SLR, a primary source was chosen~\cite{de2017api}.
Using this source as a starting point, the scope of the API-m-FAMM was determined and the initial model was constructed (\textbf{version 0.1}, Section~\ref{sec:version01}).
Subsequently, the SLR was used to populate the model, which resulted in a FAMM consisting of 114 practices and 39 capabilities that are categorized into 6 focus areas (\textbf{version 0.2}, Section~\ref{sec:version02}).
These practices and capabilities were then analyzed and verified through four validation sessions with all involved researchers, inter-rater agreement and information gathered from grey literature, such as online blog posts, websites, commercial API management platform documentation and third-party tooling  (\textbf{version 0.3}, Section~\ref{sec:version03}). 

\subsection{Test Phase}
The API-m-FAMM underwent two evaluation cycles. 
First, 11 semi-structured interviews with experts were conducted. 
During these interviews, experts were asked whether they agree with the inclusion of practices, capabilities, and focus areas as part of the API-m-FAMM, as well as whether they could suggest the addition of any new practices or capabilities. 
Additionally, practices were ranked by these experts in terms of their perceived maturity in order to determine their respective maturity levels.
As a result of these interviews, many suggestions were made to either move practices to a different capability, remove them entirely, rename them, or newly add practices. 
These suggestions were then analyzed, processed, and discussed through 6 discussion sessions with all involved researchers. 
As a result, the model was quite substantially modified, with the existing body of practices and capabilities being narrowed down to 87 practices and capabilities, as well as numerous focus areas, capabilities, and practices  being renamed. 
Additionally, all practices were assigned to individual maturity levels within their respective capabilities (\textbf{version 0.4}, Section~\ref{sec:version04}). 

The second evaluation cycle consisted of three unstructured interviews with experts originating from the sample of experts that were interviewed during the first evaluation cycle. 
During these interviews, the changes made as a result of the previous evaluation cycle, as well as the newly introduced maturity assignments were presented and discussed. 
Additionally, experts were asked to evaluate the model again with regards to the same criteria used in the first cycle. 
The API-m-FAMM was not significantly changed after this second cycle (\textbf{version 0.5}, Section~\ref{sec:version05}).

\subsection{Deploy Phase}
Finally the API-m-FAMM was used to evaluate five different software products.
The evaluation was done by using a \emph{do-it-yourself} kit, which is available on \url{https://www.movereem.nl/api-m-famm.html}.
These evaluations led to some minor changes (\textbf{version 1.0}, Section~\ref{sec:version10}).

\section{API-m-FAMM}
\label{sec:apimfamm}

The API-m-FAMM and the practices and capabilities it consists of are divided into six focus areas. The focus areas are not equal in size, with the smallest focus area consisting of 2 capabilities and 11 practices, while the largest is composed of 5 capabilities and 18 practices. This is caused by the fact that the topic of API management is broad and not evenly distributed across its domains. For example, the \textit{Community} and \textit{Lifecycle Management} focus areas that are described below contain many practices, while \textit{Observability} is a domain consisting of a small but relevant amount of practices and capabilities. 

We have defined capabilities as the ability to achieve a goal related to API Management, through the execution of two or more interrelated practices. Combined, these practices and capabilities form the focus areas which describe the functional domain the topic of API management is composed of. A practice is defined as an action that has the express goal to improve, encourage, and manage the usage of APIs. Furthermore, the practice has to be executable, implementable and verifiable by an employee of the organization.  

\begin{figure*} 
  \centering
  \includegraphics[page=1, clip, trim=0.5cm 0.5cm 0.5cm 0.5cm, width=\textwidth]{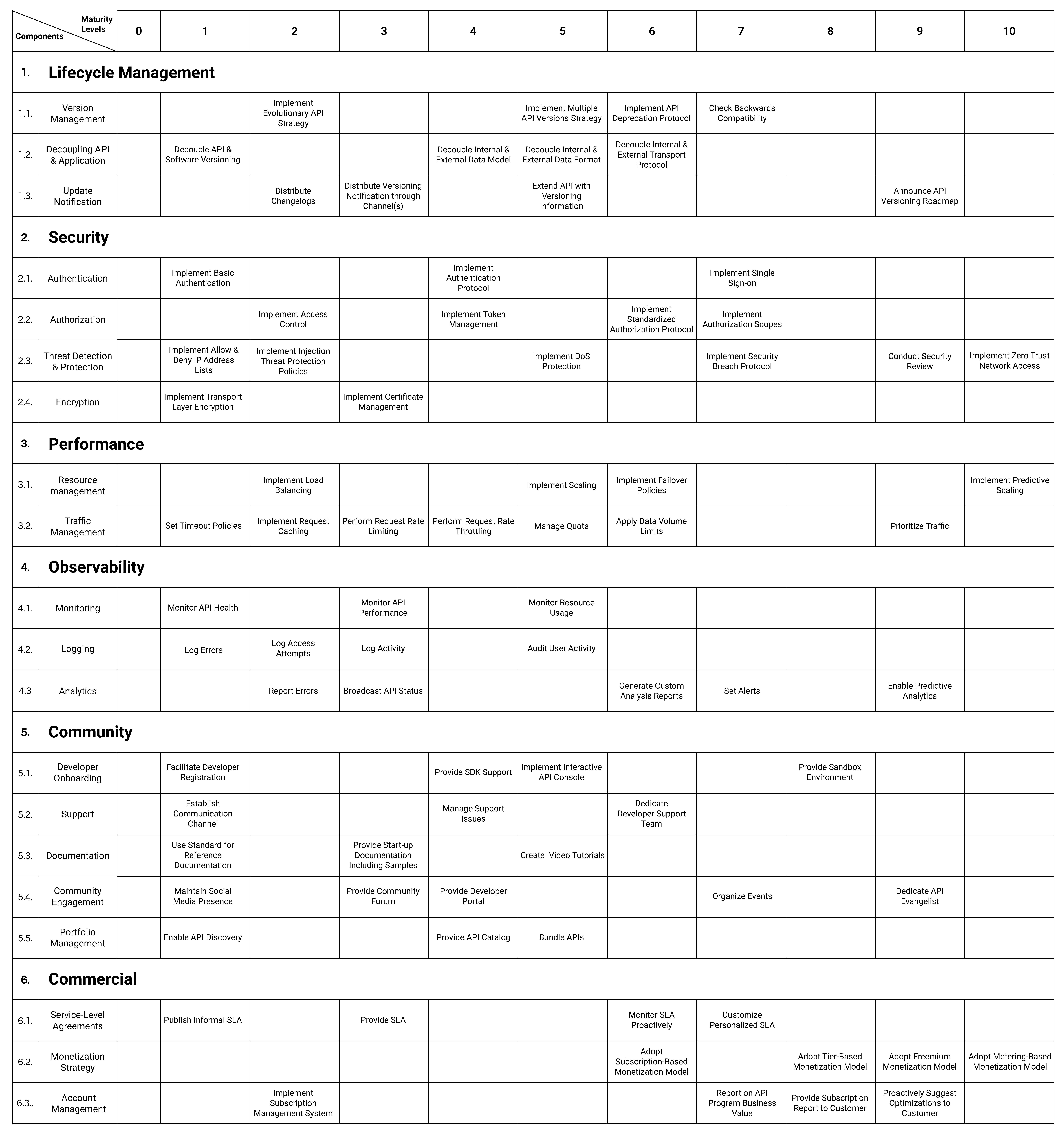}
  \caption{The API-m-FAMM model, showing all six focus areas, the capabilities, and the practices regarding API management. The columns correspond with the maturity level of the practice. }
  \label{fig:api-m-famm}
\end{figure*}

Each individual practice is assigned to a maturity level within its respective capability. As mentioned earlier, these maturity levels were determined by having experts rank the practices according to their perceived maturity within their respective capabilities. Additionally, they were asked whether they could identify any dependencies with regards to the implementation of other practices. Practices can not depend on practices as part of another capability that have a higher maturity level. For example, practice 1.1.6 is dependant on the implementation of practices 1.3.3 and 4.2.3, resulting in a higher overall maturity level being assigned to this practice. The API-m-FAMM in its entirety, including the maturity level that each practice has been assigned to, is depicted visually in Figure~\ref{fig:api-m-famm}.\\

Section~\ref{subsec:areas} describes and defines the focus areas and capabilities. Section~\ref{subsec:practices} details the practices. Practices are described by using the following elements: 

\begin{itemize}
    \item	\textbf{Practice code -} The practice code is made up of three numbers. The first number concerns the focus area, the second number the capability, and the third number the maturity level it has been assigned to.
	\item	\textbf{Practice -} The name of the practice, as it is mentioned in the API-m-FAMM.
    \item 	\textbf{Focus area -} The focus area is mentioned to indicate the domain in which this practice is relevant. 
    \item 	\textbf{Description -} A paragraph of text is provided to 
 describe the practice in detail. The main reason for providing a lengthy description is internal validity: in future evaluations by third parties, they should be able to perform the evaluations independently.
    \item 	\textbf{When implemented -} Provides a series of necessary conditions before this practice can be marked as implemented. Again, to strengthen internal validity of the API-m-FAMM.
    \item	\textbf{Literature -} Several references are included to articles that mention the practice. The literature can be found in the SLR~\cite{mathijssen2020identification}. References may also consist of online blog posts, websites, commercial API management platform documentation and third-party tooling. 
\end{itemize}

\newpage

\subsection{Focus Areas \& Capabilities}
\label{subsec:areas}

\begin{enumerate}
    \item \textbf{Lifecycle Management}: Generally speaking, an API undergoes several stages over the course of its lifetime; creation, publication, realization, maintenance and retirement \citedata{medjaoui2018continuous}. In order to control and guide the API through these stages, the organization must be able to perform a variety of activities. In order to maintain the API, the organization must decide on a versioning strategy, notification channels and methods in case of updates, as well as decouple their API from their application. In doing so, the organization is able to manage and maintain the versions the API goes through as it evolves over time.\\
     \begin{enumerate}
            \item [1.1] \textit{Version Management}: APIs evolve over time with newer business requirements. In order to cope with this, the organization should have a versioning strategy in place, such as managing multiple versions of an API to support existing consumers, or by avoiding breaking changes as part of an evolutionary strategy. Additionally, the organization should be able to deprecate and retire older versions of their API smoothly. With proper notice and period, deprecated APIs should be retired and removed so as to avoid any maintenance overheads \citedata{de2017api}. In order to guide this process, the organization may also have a deprecation protocol in place.
            \item [1.2] \textit{Decoupling API \& Application}: When an organization creates an API to expose its data and services, it needs to ensure that the API interface is intuitive enough for developers to easily use \citedata{de2017api}. However, the interface for the API will most likely be different from that of the back-end services that it exposes. Therefore, the organization should be able to transform the API interface to a form that the back end can understand.
            \item [1.3] \textit{Update Notification}: Changes made to an API may adversely affect its consumers. Hence, consumers must be notified of any planned updates of the API \citedata{de2017api}. The organization should have the ability to inform developers using the API of any changes by distributing change logs, using a communication channel such as email, the developer portal, or preemptively through the use warning headers or a versioning roadmap.\\
        \end{enumerate}
    
    \item \textbf{Security}: APIs provide access to valuable and protected data and assets \citedata{de2017api}. Therefore, security for APIs is necessary to protect the underlying assets from unauthenticated and unauthorized access. Due to the programmatic nature of APIs and their accessibility over the public cloud, they are also prone to various kinds of attacks. Hence, the organization should undertake various measures to prevent this from happening. For example, one of many available authentication and authorization protocols should be implemented, prevention for attacks such as DoS or SQL script injection attacks should be in place and sensitive data should be encrypted or masked.\\  
    \begin{enumerate}
            \item [2.1] \textit{Authentication}: Authentication is the process of uniquely determining and validating the identity of a client \citedata{de2017api}. In order to achieve this, the organization may implement an authentication mechanism such as API keys or protocols such as WSS or OpenID Connect, or the Single Sign-on method.
            \item [2.2] \textit{Authorization}: Authorization controls the level of access that is provided to an app making an API call and controls which API resources and methods that can invoke \citedata{de2017api}. The organization may implement authorization through access control or an industry-standardized authorization protocol such as OAuth 2.0.
            \item [2.3] \textit{Threat Detection \& Protection}: The likelihood of bad actors making attacks using malicious content is high, in addition to common threats such as DoS attacks. Content-based attacks can be in the form of malformed XML or JSON, malicious scripts, or SQL within the payload \citedata{de2017api}. Therefore, the organization should be able to detect malformed request formats or malicious content within the payload and then protect against such attacks.
            \item [2.4] \textit{Encryption}: Oftentimes, message payloads sent in API calls contain sensitive information that can be the target for man-in-the-middle attacks \citedata{de2017api}. Therefore, the organization should secure all communication between the client app and the API service through using techniques such as TLS encryption by default. Furthermore, it is desirable for the organization to prevent exposure of sensitive data by making utilizing methods such as masking or hashing.\\
        \end{enumerate}
    
    \item \textbf{Performance}: APIs are no longer exclusively seen as mechanisms for integration but have become mainstream for the delivery of data and services to end users through various digital channels \citedata{de2017api}. This increases the demand on APIs to perform well under loads. The overall performance of a client app is dependent on the performance of the underlying APIs powering the app. Hence, the importance of performance for APIs increases greatly. In order to ensure performance and stability of their APIs, organizations must be able to perform various activities. For example, enabling consumers to implement caching improves an API's performance through reduced latency and network traffic. Additionally, using rate limiting and throttling mechanisms to manage traffic and using load balancing to route traffic more effectively also improves the API's performance.\\
    \begin{enumerate}
            \item [3.1] \textit{Resource Management}: In order to improve the performance of their API(s), it is important for an organization to effectively manage the available resources. This may be accomplished through the use of mechanisms such as load balancing, scaling, or by having a failover policies in place.
            \item [3.2] \textit{Traffic Management}: Another aspect of improving API performance is effectively managing incoming traffic. In order to do so, the organization may choose to implement mechanisms such as caching, rate limiting or throttling, or by prioritizing traffic based on customer characteristics.\\
           \end{enumerate}
    
    \item \textbf{Observability}: As an organization, it is necessary to have insight into the API program to make the right investments and decisions during its maintenance. Through various monitoring techniques, the organization is able to collect metrics which can shed light on the API's health, performance and resource usage. In turn, these metrics may be aggregated and analyzed to improve the decision making process on how to enhance the business value by either changing the API or by enriching it \citedata{de2017api}. Additionally, by being able to log API access, consumption and performance, input may be gathered for analysis, business value or monetization reports. These may be used to strengthen communication with consumers and stakeholders or check for any potential service-level agreement violations.\\
    \begin{enumerate}
            \item [4.1] \textit{Monitoring}: As an organization, it is important to be able to collect and monitor metrics and variables concerning the exposed API. For example, information regarding the health and performance of the API, as well as resources used by the API should be monitored so that it may be used as input for activities such as generating analysis reports and broadcasting the API's operational status.
            \item [4.2] \textit{Logging}: In monitoring their API(s), it is helpful for the organization to be able to perform logging of consumer behavior and activities. This may include logging of API access, usage and reviewing historical information.
            \item [4.3] \textit{Analytics}: As an organization, it is important to be able to analyze the metrics and variables that are collected through monitoring. For example, information regarding the health and performance of the API may be utilized to decide which features should be added to the API. Additionally, it is desirable for the organization to be able to extract custom variables from within the message payload for advanced analytics reporting.\\
        \end{enumerate}

    \item \textbf{Community}: As an organization exposing APIs for external consumers and developers to consume, it is often desirable to foster, engage and support the community that exists around the API. For example, this entails offering developers the ability register on the API and offering them access to test environments, code samples and documentation. Additionally, the organization may support developers in their usage of the API by offering them support through a variety of communication channels and allowing them to communicate with the organization or among another through a community forum or developer portal. Furthermore, it is desirable for developers to be able to freely browse through the API offering, review operational status updates regarding the API, create support tickets in the event of an error and to share knowledge, views and opinions with other developers.\\ 
    \begin{enumerate}
            \item [5.1] \textit{Developer Onboarding}: To start consuming APIs, developers must first register with the organization that is providing them. The sign up process should be simple and easy, possibly by supporting developers with resources such as (automatically generated) SDKs and testing tools such as an API console or sandbox environment.
            \item [5.2] \textit{Support}: In order to strengthen the community around the API, the organization should support developers whom are consuming it. This may be accomplished by establishing an appropriate communication channel, adequately managing issues and handling errors, should they present themselves.
            \item [5.3] \textit{Documentation}: API documentation can help speed up the adoption, understanding and effectiveness of APIs \citedata{de2017api}. Hence, the organization must provide consumers of their API(s) with reference documentation. Additionally, they may be supplied with start-up documentation, code samples and FAQs to further accelerate understanding of the API.
            \item [5.4] \textit{Community Management}: Oftentimes, app developers wish to know the views of other developers in the community. They may want to collaborate and share their API usage learnings and experiences with one another \citedata{de2017api}. In order to facilitate these wishes, the organization may choose to provide developers with a community forum or developer portal.
            \item [5.5] \textit{Portfolio Management}: As an API providing organization, a platform to publicize and document APIs is needed. Hence, a discoverable catalog of APIs through which potential consumers are able to browse may be provided.\\
        \end{enumerate}
    
    \item \textbf{Commercial}: Organizations have been consuming third-party APIs to simplify and expand business partnership. APIs provide faster integration and an improved partner/customer experience, enabling organizations to grow rapidly \citedata{de2017api}. Oftentimes, exposing and consuming APIs has a commercial aspect tied to it. For API consumers and providers, this is often embodied by legal business contracts for the use of the APIs which they are bound to. These business contracts called service-level agreements govern the service levels and other aspects of API delivery and consumption. Another commercial aspect of API management is that of monetization. Considering APIs provide value to the consuming party, organizations often opt to monetize the services and APIs and build a business model for them \citedata{de2017api}. Utilizing the right monetization model for APIs enables organizations to reap the benefits of their investment in their APIs.\\
     \begin{enumerate}
        \item [6.1] \textit{Service-Level Agreements}: A service-level  agreement (SLA) defines the API’s  non-functional  requirements, serving as a contract between the organization and consumers of their API. As such, the organization should ensure that the consumer of their API agrees with the SLA's contents. These may include matters such as terms and conditions for API usage, consumption quotas, uptime guarantees and  maintenance or downtime information.
        \item [6.2] \textit{Monetization Strategy}: APIs securely expose digital assets and services that are of value to consumers. Hence, the organization may wish to adopt a monetization strategy to enable monetization of the exposed services and APIs by constructing a business model around them. This may be accomplished through a monetization model which can be based on consumer characteristics such as their type of subscription, access tier or the amount of resources used.
        \item [6.3] \textit{Account Management}: It is desirable to effectively manage accounts in order to foster a qualitative relationship with customers, stakeholders and the organization's management. This may be achieved by reporting on the API's business value internally through the use of business value reports, as well as externally by providing consumers of the API with subscription reports and training them in using the API as efficiently as possible. \\
      \end{enumerate}
\end{enumerate}

\subsection{Practices}
\label{subsec:practices}

\newarray\MyData
\readarray{MyData}
{
1.1.2 &
Implement Evolutionary API Strategy &
Version Management &
Lifecycle Management &
The organization utilizes an evolutionary strategy to continuously version their API over time. Using this strategy, the organization evolves a single API by avoiding the introduction of breaking changes. Optionally, this may be accomplished by adhering to the GraphQL specification \citedata{graphqlVersioning}. &
$\bullet$ The organization maintains one version of their API. \newline
$\bullet$ The organization utilizes an evolutionary API versioning strategy.
& \citedata{ploesserVersioning, icappsVersioning}     &
 &
6& 
1.1.5 &
Implement Multiple API Versioning Strategy &
Version Management &
Lifecycle Management &
The organization has a versioning strategy in place which entails the process of versioning from one API to a newer version. In order to do so, the organization must be able to maintain multiple versions of (one of) their API(s) for a period of time. Possible strategies include URI/URL Versioning (possibly in combination with adherence to the Semantic Versioning specification), Query Parameter versioning, (Custom) Header versioning, Accept Header versioning or Content Negotiation.  &
$\bullet$ The organization utilizes one of the following versioning strategies: URI/URL Versioning, Query Parameter versioning, (Custom) Header versioning, Accept Header versioning or Content Negotiation.
& \citedata{de2017api, redhatVersioning, anjiVersioning, rapidVersioning}      &
 &
6& 
1.1.6 &
Implement API Deprecation Protocol &
Version Management &
Lifecycle Management &
The organization has a protocol in place that details what steps should be taken when deprecating one of their APIs. This includes determining the amount of developers currently consuming the API through the use of monitoring, and then setting a threshold that details the amount of developers that should have migrated to the new version of the API before commencing with deprecation of the old version. Furthermore, developers, including their contact information, should be identified so that they may be notified of the deprecation through their preferred communication channel. This notification should be accompanied by a migration period and deprecation date, so that consumers have a clear target to migrate their apps over to the new API version. Additionally, referrals to to documentation and the new endpoint should be included. Furthermore, the protocol should detail what course of action should be taken to roll back to a previously deployed version of an API in the event of an incorrect deployment of the API. &
$\bullet$ The organization has implemented the 'Distribute Versioning Notification Through Channel(s)' (1.3.3) and 'Log Activity' (4.2.3) practices. \newline
$\bullet$ The organization has a deprecation protocol in place.
&  \citedata{peterLifecycle} &
 &
6& 
1.1.7 &
Check Backwards Compatibility &
Version Management &
Lifecycle Management &
The organization has an approach in place with which it is able to detect breaking changes when versioning their API(s). Approaches include using a unit test suite, plugging an automated contract test suite into the CI/CD pipeline or by using the \emph{swagger-spec-compatibility} library to detect differences between two Swagger / OpenAPI specifications \citedata{swaggerComp}. &
$\bullet$ The organization has implemented the 'Implement Evolutionary API Versioning Strategy' (1.1.2) practice. \newline
$\bullet$ The organization has a backwards compatibility checking approach in place.
& \citedata{bhojwaniCheck}      &
 &
6& 
1.2.1 &
Decouple API \& Software Versioning &
Decoupling API \& Application &
Lifecycle Management &
The organization has decoupled the version of their API(s) from its software implementation. The API version should never be tied to the software version of the back-end data/service. A new API version should be created only if there is a change in the contract of the API that impacts the consumer. &
$\bullet$ The organization has decoupled the version of their API(s) from its software implementation.
& \citedata{de2017api}  &
 &
6& 
1.2.4 &
Decouple Internal \& External Data Model &
Decoupling API \& Application &
Lifecycle Management &
The organization has decoupled the data models that are used internally and externally from one another. Doing so is considered to be beneficial, since an application might use a normalized relation data model internally. While this data model is less suitable to expose through a public API, this separation of concerns allows the organization to evolve the relational data model at a different speed than the API.
& The organization has decoupled the data models that are used internally and externally from one another & None. &
 &
6& 
1.2.5 &
Decouple Internal \& External Data Format
&
Decoupling API \& Application &
Lifecycle Management &
The organization has decoupled the data format that are used internally and externally from one another. Doing so is considered to be beneficial, since an application might use a data format such as XML internally, while using a data format such as JSON for the API(s). This separation of concerns grants the organization greater flexibility in designing and developing their APIs.
&
$\bullet$ The organization has decoupled the data format that are used internally and externally from one another.
&  None.  &
 &
6& 
1.2.6 &
Decouple Internal \& External Transport Protocol &
Decoupling API \& Application &
Lifecycle Management &
The organization has decoupled the transport protocol that are used internally and externally from one another. Considering that an application might internally use a protocol that is less commonly used in modern APIs such as SOAP or JDBC internally, which may be less suitable for public APIs, the organization may opt to use a different protocol for their API(s). This separation of concerns grants the 
These protocols are less commonly used in modern APIs, or are less suitable for public APIs, and the organization can decide to use a different protocol for the APIs. This separation of concerns grants the organization greater flexibility in designing and developing their APIs.
&
$\bullet$ The organization has decoupled the transport protocol that are used internally and externally from one another.
& None.   &
 &
6&
1.3.2 &
Distribute Changelogs &
Update Notification &
Lifecycle Management &
The organization uses (automated) email services to distribute changelogs describing the versioning of their API(s) to consumers. Ideally, the organization offers consumers the ability to opt-in or opt-out of this service. &
$\bullet$ The organization uses (automated) email services to distribute changelogs describing the versioning of their API(s) to consumers. & \citedata{sandovalChange} &
 &
6& 
1.3.3 &
Distribute Versioning Notification Through Channel(s) &
Update Notification &
Lifecycle Management &
The organization has the ability to distribute versioning notifications among consumers of their API(s) through established communication channels. Possible channels include email, social media, and announcements within the developer portal or reference documentation. Ideally, the organization offers consumers of their API(s) the option to select the communication channel they prefer receiving versioning notifications through.
&
$\bullet$ The organization has implemented the 'Establish Communication Channel' (5.2.1) and 'Distribute Changelogs' (1.3.2) practices. \newline
$\bullet$ The organization has the ability to distribute versioning notifications among consumers of their API(s) through established communication channels. 
&  \citedata{de2017api, sandovalChange}    &
  &
6& 
1.3.5 &
Extend API with Versioning Information &
Update Notification &
Lifecycle Management &
The organization has the ability to extend their API specification to incorporate warning headers into responses in run-time. By doing so, consumers of the API are notified of its impending deprecation, and possibly requested to change their implementation. &
$\bullet$ The organization has the ability to introduce warning headers.
& \citedata{de2017api} &
  &
6& 
1.3.9 &
Announce Versioning Roadmap &
Update Notification &
Lifecycle Management &
The organization has announced a roadmap that details the planned dates on which the current (old) version of their API will be versioned to a new version, in order to notify consumers ahead of time. This may be done through email, social media, announcements within the developer portal or reference documentation.&
$\bullet$ The organization has implemented the 'Distribute Versioning Notification Through Channel(s)' (1.3.3) practice. \newline
$\bullet$ The organization has announced a versioning roadmap.
&   \citedata{de2017api}   &
  &
6& 
2.1.1 &
Implement Basic Authentication &
Authentication &
Security &
The organization has the ability to implement basic authentication in order to authenticate consumers of their API(s). This may be accomplished through the use of HTTP Basic Authentication, with which the consumer is required to provide a username and password to authenticate, or by issuing API keys to consumers of the API. An app is identified by its name and a unique UUID known as the API key, often serving as an identity for the app making a call to the API. &
$\bullet$ The organization has implemented HTTP Basic Authentication, or is able to issue API keys.
&  \citedata{biehl2015api, de2017api, Zhao_2018, sandoval2018_2}    &
 &
6& 
2.1.4 &
Implement Authentication Protocol &
Authentication &
Security &
The organization has implemented an authentication protocol or method in order to authenticate consumers of their API(s). In order to apply security For SOAP APIs, the usage of a WS Security (WSS) protocol \citedata{wikipediaWS} may be opted for. This protocol specifies how integrity and confidentiality can be enforced on messages and allows the communication of various security token formats, such as Security Assertion Markup Language (SAML), X.509 and User ID/Password credentials. Consumers of REST APIs may be authenticated by using methods and protocols such as Client Certificate authentication, SAML authentication, or OpenID Connect \citedata{openIDConnect}. OpenID Connect 1.0 is an authentication protocol that builds on top of OAuth 2.0 specs to add an identity layer. It extends the authorization framework provided by OAuth 2.0 to implement authentication.&
$\bullet$ The organization has implemented a WSS authentication protocol, or methods and protocols such as Client Certificate authentication, SAML authentication, or OpenID Connect.
&  \citedata{de2017api, oracleWS, wikipediaWS}   &
 &
6& 
2.1.7 &
Implement Single Sign-On &
Authentication &
Security &
The organization has implemented Single Sign-on (SSO), which is a authentication method that enables users to securely authenticate with multiple applications and websites by using one set of credentials. The user is then signed in to other applications automatically, regardless of the platform, technology, or domain the user is using. 
&
$\bullet$ The organization has implemented the 'Implement Authentication Protocol' (2.1.4) practice. \newline
$\bullet$ The organization has implemented the Single Sign-on (SSO) authentication method.
&  \citedata{de2017api, Onelogin, SSO}    &
 &
6& 
2.2.2 &
Implement Access Control &
Authorization &
Security &
The organization has implemented an access control method in order to identify and authorize consumer potential users of their API(s). In order to accomplish this, the Role-based Access Control (RBAC) method may be used, with which permissions may be assigned to users based on their role within the organization. Alternatively, the Attribute-based Access Control (ABAC) may be used, with which permissions are granted based on an identities' attributes. Optionally, RBAC and ABAC policies may be expressed by using the eXtensible Access Control Markup Language (XACML).
&
$\bullet$ The organization has implemented the Role-based Access Control (RBAC) or Attribute-based Access Control (ABAC) method.
&  \citedata{de2017api, hofman2014technical, thielens2013apis, WikiXACML}   &
 &
6& 
2.2.4 &
Implement Token Management &
Authorization &
Security &
The organization provides consumers of their API(s) with the ability to perform (access) token and API key management. This is an activity that involves measures to manage (i.e. review, store, create and delete) the tokens and API keys that are required to invoke back-end APIs. &
$\bullet$ The organization allows consumers to manage their tokens and API keys.
&  \citedata{de2017api, hofman2014technical}    &
 &
6& 
2.2.6 &
Implement Standardized Authorization Protocol &
Authorization &
Security &
The organization has implemented an industry-standardized authorization protocol, such as the OAuth 2.0 Authorization protocol. OAuth is used as a mechanism to provide authorization to a third-party
application for access to an end user resource on behalf of them. OAuth helps with granting authorization without the need to share user credentials. &
$\bullet$ The organization has an industry-standardized authorization protocol. 
&   \citedata{de2017api,gadge2018microservice,gamez2015towards,hohenstein2018architectural,matsumoto2017fujitsu,patni2017pro,thielens2013apis,hofman2014technical,Xu_2019,Zhao_2018}  &
 &
6& 
2.2.7 &
Implement Authorization Scopes &
Authorization &
Security &
The organization has implemented an authorization scopes mechanism, such as the OAuth 2.0 Scopes mechanism \citedata{OAuthScopes}, to limit access to their application(s) to their users' accounts. An application can request one or more scopes, where after this information is then presented to the user in a consent screen. Then, the access token that was issued to the application will be limited to the scopes granted.  &
$\bullet$ The organization has an authorization scopes mechanism in place. 
&   None.  &
 &
6& 
2.3.1 &
Implement Allow \& Deny IP Address Lists &
Threat Detection \& Protection &
Security &
The organization has the ability to impose allow and deny list policies. Through these policies, specific IPs can either be excluded from requests, or separate quotas can be given to internal users by throttling access depending on their IP address or address range.
&
$\bullet$ The organization has the ability to impose allow and deny list policies.
&  \citedata{gadge2018microservice, gamez2015towards, hohenstein2018architectural}    &
 &
6& 
2.3.2  &
Implement Injection Threat Protection Policies &
Threat Detection \& Protection &
Security &
The organization has implemented injection threat protection security policies. Injection threats are common forms of attacks, in which attackers try to inject malicious code that, if executed on the server, can divulge sensitive information. These attacks may take the form of XML and JSON bombs or SQL and script injection.&
$\bullet$ The organization has injection threat policies in place against XML or JSON bombs or SQL or script injection.
&  \citedata{de2017api, preibisch2018api, OWASPInjection}    &
 &
6& 
2.3.5 &
Implement DoS Protection &
Threat Detection \& Protection &
Security &
The organization has protection against DoS attacks in place. Hackers may try to bring down back-end systems by pumping unexpectedly high traffic through the APIs. Denial-of-service (DoS) attacks are very common on APIs. Hence, the organization should be able to detect and stop such attacks. Identification of a DoS attack is done through Spike Arrest. &
$\bullet$ The organization has protection against DoS attacks in place.
&  \citedata{de2017api, gadge2018microservice, gamez2015towards}      &
  &
6& 
2.3.7  &
Implement Security Breach Protocol &
Threat Detection \& Protection &
Security &
The organization has a security breach protocol in place, which details what steps should be taken in the event where a security breach occurs. This protocol may include activities such as notifying stakeholders and consumers of the API, identifying the source of the breach by scanning activity logs, containing the breach by stopping the data leakage, and consulting third-party IT security and legal advice providers.
&
$\bullet$ The organization has a security breach protocol in place.
&  \citedata{Reynold2020, Soliya2020}    &
 &
6& 
2.3.9  &
Conduct Security Review &
Threat Detection \& Protection &
Security &
The organization has the ability to conduct security reviews that potential consumers of their API(s) must pass before being allowed to integrate the organization's API(s) into their application. This typically involves testing the degree to which customer data is protected and encrypted, and identifying security vulnerabilities that may be exploited, such as threats related to script injections and non-secure authentication and access control protocols. 
&
$\bullet$ The organization has the ability to conduct security reviews.
&  \citedata{Salesforce2020} &
 &
6& 
2.3.10  &
Implement Zero Trust Network Access (ZTNA) &
Threat Detection \& Protection &
Security &
The organization has implemented a Zero Trust Network Access (ZTNA) security architecture, where only traffic from authenticated users, devices, and applications is granted access to other users, devices, and applications within an organization. ZTNA may be regarded as a fine-grained approach to network access control (NAC), identity access management (IAM) and privilege access management (PAM), offering a replacement for VPN architectures. Optionally, a ZTNA may be implemented through third-party providers such as Akamai, Cloudflare, or Cisco. 
&
$\bullet$ The organization has implemented a Zero Trust Network Access (ZTNA) security architecture.
&  \citedata{ZTNAwiki2020}    &
 &
6& 
2.4.1 &
Implement Transport Layer Encryption &
Encryption &
Security &
The organization has implemented current and up-to-date encryption protocols such as Transport Layer Security (TLS). It is always desirable to have TLS compliant endpoints to safeguard against man-in-middle attacks, and bi-directional encryption of message data to protect against tampering. &
$\bullet$ The organization has implemented a current and up-to-date transport layer encryption protocol. 
&  \citedata{de2017api, familiar2015iot, gadge2018microservice, hofman2014technical, preibisch2018api}    &
 &
6& 
2.4.3 &
Implement Certificate Management &
Encryption &
Security &
The organization has the ability to manage its TLS certificates. This involves monitoring and managing the certificates' acquisition and deployment, tracking renewal, usage, and expiration of SSL/TLS certificates. &
$\bullet$ The organization has the ability to manage its TLS certificates.
& \citedata{de2017api,hohenstein2018architectural,sine2015api,thielens2013apis,gadge2018microservice}     &
 &
6& 
3.1.2 &
Implement Load Balancing &
Resource Management &
Performance &
The organization has implemented load balancing to distribute API traffic to the back-end services. Various load balancing algorithms may be supported. Based on the selected algorithm, the requests must be routed to the appropriate resource that is hosting the API. Load balancing also improves the overall performance of the API. &
$\bullet$ The organization has implemented load balancing.
& \citedata{biehl2015api,ciavotta2017microservice,de2017api,gadge2018microservice,gamez2015towards,montesi2016circuit,nakamura2017fujitsu,Xu_2019,Zhao_2018} &
 &
6& 
3.1.5 &
Implement Scaling  &
Resource Management &
Performance &
The organization has the ability to scale the amount of available resources up or down depending on traffic and API usage in a reactive manner. This may be done either manually or automatically, through the use of a load balancer. &
$\bullet$ The organization has implemented the 'Implement Load Balancing' (3.1.2) practice. \newline
$\bullet$ The organization has the ability to scale the amount of available resources up or down.
&  \citedata{akbulut2019software,jacobson2011apis,gadge2018microservice,hofman2014technical}     &
  &
6& 
3.1.6 &
Implement Failover Policies &
Resource Management &
Performance &
The organization has the ability to mitigate outages through the implementation of failover policies. This may be done by automatically deploying a service to a standby data center if the primary system fails, or is shut down for servicing. By being able to perform a failover, the particular service is guaranteed to be operational at one of the data centers. This is an extremely important function for critical systems that require always-on accessibility.  &
$\bullet$ The organization has the ability to mitigate outages through the implementation of failover policies.
&  \citedata{Barracuda2020}  &
   &
6& 
3.1.10 &
Implement Predictive Scaling  &
Resource Management &
Performance &
The organization has the ability to scale the amount of available resources up or down depending on traffic and API usage in a proactive manner. This may be done automatically, through the use of a load balancer as based on insights gained from predictive analytics. &
$\bullet$ The organization has implemented the 'Implement Load Balancing' (3.1.2) and 'Enable Predictive Analytics' (4.3.9) practices. \newline
$\bullet$ The organization has implemented predictive scaling.
&  None. &
  &
6& 
3.2.1 &
Set Timeout Policies &
Traffic Management &
Performance &
The organization is able to set timeout policies, by detecting and customizing the amount of time that is allowed to pass before a connection times out and is closed. Using timeout policies, the organization is able to ensure that the API always responds within a given amount of time, even if a long-running process hangs. This is important in high-availability systems where response performance is crucial so errors can be dealt with cleanly. &
$\bullet$ The organization is able to set timeout policies on their API(s).
&  \citedata{tykTimeout}     &
    &
6& 
3.2.2 &
Implement Request Caching &
Traffic Management &
Performance &
The organization utilizes caching as a mechanism to optimize performance. As consumers of the API make requests on the same URI, the cached response can be used to respond instead of forwarding those requests to the back-end server. Thus caching can help to improve an APIs performance through reduced latency and network traffic. &
$\bullet$ The organization utilizes caching as a mechanism to optimize performance.
& \citedata{biehl2015api,de2017api,gadge2018microservice,gamez2015towards,indrasiri2018developing,patni2017pro,preibisch2018api,vsnuderl2018rate,vijayakumar2018practical,hofman2014technical,Zhao_2018}     &
  &
6& 
3.2.3 &
Perform Request Rate Limiting &
Traffic Management &
Performance &
The organization has a mechanism in place with which limits on the amount of requests or faulty calls API consumers are allowed to make, may be imposed. Requests made within the specified limit are routed successfully to the target system. Those beyond the limit are rejected. &
$\bullet$ The organization has a rate limiting mechanism in place for their API(s).
&   \citedata{de2017api,gamez2015towards,jacobson2011apis,lourencco2019framework,raivio2011towards,jayathilaka2015eager,vsnuderl2018rate,hofman2014technical,gadge2018microservice}     &
  &
6& 
3.2.4 &
Perform Request Rate Throttling &
Traffic Management &
Performance &
The organization has a mechanism in place with which API requests may be throttled down, without the connection being closed. This can help to improve the overall performance and reduce impacts during peak hours. It helps to ensure that the API infrastructure is not slowed down by high volumes of requests from a certain group of customers or apps. &
$\bullet$ The organization has a rate throttling mechanism in place for their API(s).
& \citedata{de2017api,fremantle2015web,familiar2015iot,gadge2018microservice,hohenstein2018architectural,indrasiri2018developing,jacobson2011apis,thielens2013apis,weir2015oracle}  &
  &
6& 
3.2.5 &
Manage Quota &
Traffic Management
&
Performance &
The organization has policies in place regarding the number of API calls that an app is allowed to make to the back end over a given time interval. Calls exceeding the quota limit may be throttled or halted. The quota allowed for an app depends on the business policy and monetization model of the API. A common purpose for a quota is to divide developers into categories, each of which has a different quota and thus a different relationship with the API. &
$\bullet$ The organization has implemented the 'Perform Request Rate Limiting' (3.2.3) practice or 'Perform Request Rate Throttling' (3.2.4) practice.\newline
$\bullet$ The organization has quota policies for their API(s) in place.
&  \citedata{de2017api}       &
 &
6& 
3.2.6 &
Apply Data Volume Limits &
Traffic Management &
Performance &
The organization has a mechanism in place with which the amount of data consumers of their API(s) are allowed to consume in one call may be limited. This can help to improve the overall performance and reduce impacts during peak hours. It helps to ensure that the API infrastructure is not slowed down by calls that transport unnecessarily high chunks of data volumes. &
$\bullet$ The organization has implemented the 'Monitor Resource Usage' (4.1.5) practice.\newline
$\bullet$ The organization has a data volume limiting mechanism in place.
& \citedata{DropboxDatalimiting}  &
  &
6& 
3.2.9 &
Prioritize Traffic &
Traffic Management &
Performance &
The organization is able to give a higher priority in terms of processing API calls, based on certain customer characteristics and/or classes. This priority may be based on their subscription, customer relationships, or agreements made in the SLA. &
$\bullet$ The organization is able to prioritize traffic based on customer characteristics and/classes.
&\citedata{de2017api}      &
  &
6& 
4.1.1 &
Monitor API Health  &
Monitoring &
Observability &
The organization is able to perform health monitoring on its API(s), possibly through an management platform, external monitoring tool/dashboard, functional testing or custom scripts and plugins. This should return basic information such as the operational status of the API, indicating its ability to connect to dependent services.  &
$\bullet$ The organization is able to perform health monitoring on its API(s).
& \citedata{averdunkHealth, gadge2018microservice}      &
 &
6& 
4.1.3 &
Monitor API Performance  &
Monitoring &
Observability &
The organization is able to perform performance monitoring on its API(s), possibly through an management platform, external monitoring tool/dashboard, functional testing or custom scripts and plugins. Doing so should provide performance statistics that track the latency within the platform and the latency for back-end calls. This helps the organization in finding the source of any performance issues reported on any API. &
$\bullet$ The organization is able to perform performance monitoring on its API(s).
&  \citedata{de2017api, Xu_2019}    &
 &
6& 
4.1.5 &
Monitor Resource Usage  &
Monitoring &
Observability &
The organization is able to perform resource monitoring on its API(s), possibly through an management platform, external monitoring tool/dashboard, functional testing or custom scripts and plugins. Doing so should provide insights into the amount of resources that are consumed as a result of calls made to the API(s). This may be done by measuring 
hardware metrics such as CPU, disk, memory, and network usage, or by using an indirect approximation of the amount of resources that are consumed by calls. &
$\bullet$ The organization is able to perform resource monitoring on its API(s).
&  \citedata{KubernetesResources}  &
 &
6& 
4.2.1 &
Log Errors &
Logging &
Observability &
The organization has the ability to internally log errors that are generated as a result of consumption of their APIs. Error logs should typically contain fields that capture information such as the date and time the error has occurred, the error code, and the client IP and port numbers.
&
$\bullet$ The organization has the ability to internally log errors.
&    \citedata{andrey_kolychev_konstantin_zaytsev_2019_3256462, de2017api, medjaoui2018continuous} &
 &
6& 
4.2.2 &
Log Access Attempts &
Logging &
Observability &
The organization has the ability to generate access logs, in which HTTP requests/responses are logged, to monitor the activities related to an APIs usage. Access logs offer insight into who has accessed the API, by including information such as the consumer's IP address. &
$\bullet$ The organization is able to perform access logging.
& \citedata{wso2Access}   &
 &
6& 
4.2.3 &
Log Activity &
Logging &
Observability &
The organization has the ability to perform basic logging of API activity, such as access, consumption, performance, and any exceptions. In doing so, it may be determined what initiated various actions to allow for troubleshooting any errors that occur. &
$\bullet$ The organization is able to perform activity logging.
&   \citedata{de2017api, fremantle2015web, gadge2018microservice} &
 &
6& 
4.2.5 &
Audit User Activity  &
Logging &
Observability &
The organization is able to perform user auditing. Doing so enables the organization to review historical information regarding API activity, to analyze who accesses an API, when it is accessed, how it is used, and how many calls are made from the various consumers of the API. &
$\bullet$ The organization is able to perform user auditing.
& \citedata{de2017api, gadge2018microservice}      &
 &
6& 
4.3.2 &
Report Errors  &
Analytics &
Observability &
The organization has the ability to report any errors to consumers that may occur during usage of their API(s). Error reports typically include information such as the error code and text describing why the error has occurred. &
$\bullet$ The organization has implemented the 'Log Errors' (4.2.1) practice.\newline
$\bullet$ The organization is able to report any errors to consumers.
& \citedata{andrey_kolychev_konstantin_zaytsev_2019_3256462, de2017api, medjaoui2018continuous}      &
 &
6& 
4.3.3 &
Broadcast API Status  &
Analytics &
Observability &
The organization broadcasts the status of its API(s) to consumers by providing them with operational information on the API in the form of an external status page, possibly on the developer portal or a website. The function of this status page is to let consumers know what is going on with the API at a technical level at any point in time. &
$\bullet$ The organization has implemented the 'Monitor API Health' (4.1.1) practice.\newline
$\bullet$ The organization broadcasts the operational status of its API(s) to consumers.
& \citedata{sandoval2018}      &
  &
6& 
4.3.6 &
Generate Custom Analysis Reports &
Analytics &
Observability &
The organization is able to generate custom analysis reports on metrics of choice, possibly through an API management platform or monitoring tool. &
$\bullet$ The organization is able to generate custom analysis reports.
& \citedata{de2017api}     &
 &
6& 
4.3.7 &
Set Alerts  &
Analytics &
Observability &
The organization has the ability to set and configure alerts that should trigger in case of certain events or thresholds being exceeded. Such events or thresholds may include resource limits being exceeded, or occurrence of outages. Ideally, the organization is able to configure what persons should be alerted about the event, and through what communication channel they should be contacted.   &
$\bullet$ The organization has implemented the 'Monitor API Health' (4.1.1), 'Monitor API Performance' (4.1.3), and 'Monitor API Resource Usage' (4.1.5) practices.\newline
$\bullet$ The organization has the ability to set and configure alerts.
& \citedata{UptrendsAlerting}    &
 &
6& 
4.3.9 &
Enable Predictive Analytics  &
Analytics &
Observability &
The organization has the ability to aggregate predictive analytics, through techniques such as pattern recognition, data mining, predictive modelling, or machine learning, by analyzing current and historical facts to make predictions about future or otherwise unknown events.   &
$\bullet$ The organization has implemented the 'Monitor API Performance' (4.1.3) and 'Monitor API Resource Usage' (4.1.5) practices.\newline
$\bullet$ The organization has the ability to aggregate predictive analytics.
& None.  &
 &
6& 
5.1.1 & 
Facilitate Developer Registration & 
Developer Onboarding &
Community & 
The organization has a mechanism in place with which API consumers are able to register to the API so that they can obtain access credentials. Consumers can then select an API and register their apps to use it. &
$\bullet$ The organization has a mechanism in place with which API consumers are able to register to their API(s). &
\citedata{de2017api} & 
 &
6&
5.1.4 &
Provide SDK Support &
Developer Onboarding &
Community &
The organization offers API consumers the option to either download client-side SDKs for the API, or generate the SDK themselves from standard API definition formats such as OpenAPI (formerly known as Swagger). These functionalities are usually offered through the developer portal, where app developers often look for device-specific libraries to interact with the services exposed by the API. &
$\bullet$ The organization offers API consumers the option to download or generate client-side SDKs for their API(s).
& 
\citedata{de2017api} &
  &
6&
5.1.5 &
Implement Interactive API Console &
Developer Onboarding &
Community &
The organization provides API consumers with an interactive console. Using this console, developers are able to test the behavior of an API. &
$\bullet$ The organization provides API consumers with an interactive console.  &
\citedata{biehl2015api} &
   &
6&
5.1.8 &
Provide Sandbox Environment Support &
Developer Onboarding &
Community &
The organization provides API consumers with an environment that they can use to mimic the characteristics of the production environment and create simulated responses from all APIs the application relies on. &
$\bullet$ The organization provides API consumers with a sandbox environment.
&
\citedata{buidesign, jacobson2011apis, Mueller:2020, patni2017pro} & 
 &
6& 
5.2.1 &
Establish Communication Channel &
Support &
Community &
The organization has established a communication channel between the API provider and consumer with which support may be provided to the consumer. Possible communication media include email, phone, form, web, community forum, blogs or the developer portal.&
$\bullet$ The organization has established one of the following communication channels with consumers of their API(s): email/phone/form/web/ community forum/blog/developer portal. & 
\citedata{de2017api, jacobson2011apis}     &
   &
6 & 
5.2.4 &
Manage Support Issues &
Support &
Community &
The organization is able to manage any support issues with their API(s). API consumers must be able to report any issues, bugs or shortcomings related to the API. They should be able to raise support tickets and seek help regarding API usage. Additionally, the API provider must be able to track and prioritize support tickets. &
$\bullet$ The organization is able to manage any support issues with their API(s).
&  \citedata{de2017api, jacobson2011apis}     &
   &
6& 
5.2.6 &
Dedicate Developer Support Team  &
Support &
Community &
The organization employs a dedicated  that offers support to consumers of their API(s). This team should be well-trained and possess knowledge that enables them to assist consumers with any problems or difficulties they may experience during the usage or implementation of the API. &
$\bullet$ The organization has implemented the 'Establish Communication Channel' (5.2.1) practice. \newline
$\bullet$ The organization employs a dedicated developer team that offers support to consumers of their API(s).
& None. &
 &
6& 
5.3.1 &
Use Standard for Reference Documentation &
Documentation &
Community &
The organization provides consumers of their API(s) with basic reference documentation on their website, developer portal or an external, third-party documentation platform. This documentation should document every API call, every parameter, and every result so that consumers are informed on the API's functionality. Additionally, it must be specified using a documentation framework such as Swagger, RAML, API Blueprint, WADL, Mashery ioDocs, Doxygen, ASP.NET API Explorer, Apigee Console To-Go, Enunciate, Miredot, Dexy, Docco or TurnAPI. &
$\bullet$ The organization provides consumers of their API(s) with basic reference documentation.\newline 
$\bullet$ The organization utilizes one of the following (or comparable) documentation tools to specify its API documentation: Swagger (OpenAPI), RAML, API Blueprint, WADL, Mashery ioDocs, Doxygen, ASP.NET API Explorer, Apigee Console To-Go, Enunciate, Miredot, Dexy, Docco or TurnAPI.
&  \citedata{de2017api, jacobson2011apis, medjaoui2018continuous}   &
 &
6&  
5.3.3 & 
Provide Start-up Documentation \& Code Samples &
Documentation &
Community &
The organization provides consumers of their API(s) with start-up documentation on on their website, developer portal or an external, third-party documentation platform. This type of documentation explains key concepts by summarizing the reference documentation, accelerating understanding as a result. Optionally, a list of Frequently Asked Questions and code samples that may be readily used in apps to invoke the API may be included.
&
$\bullet$ The organization has implemented the 'Use Standard for Reference Documentation' (5.3.1) practice. \newline
$\bullet$ The organization provides consumers of their API(s) with start-up documentation.
& \citedata{de2017api, jacobson2011apis}  &
    &
6& 
5.3.5 &
Create Video Tutorials &
Documentation &
Community &
The organization is able to create video tutorials in order to provide consumers with visual information that details how to use the API and integrate it into their applications.
&
$\bullet$ The organization is able to create video tutorials.
& None.   &
  &
6& 
5.4.1 &
Maintain Social Media Presence &
Community Engagement &
Community  &
The organization is able to maintain their social media presence on platforms such as Facebook or Twitter. This may involve activities such as reporting on the API's status, announcing news and updates, responding to questions, or reacting to feedback.
&
$\bullet$ The organization is able to maintain their social media presence on platforms such as Facebook or Twitter.
&   None. &
 &
6& 
5.4.3 &
Provide Community Forum &
Community Engagement &
Community  &
The organization provides (potential) consumers of their API(s) with a community forum, possibly through a website or API management platform. This forum may assist in building and interconnecting a developer community, by providing them with a central hub they can use to communicate with one another and the organization. Additionally, it may serve as a repository with guides on API usage, documentation and support. &
$\bullet$ The organization provides API consumers with a community forum.
&  \citedata{de2017api}   &
  &
6& 
5.4.4 &
Provide Developer Portal &
Community Engagement &
Community &
The organization provides (potential) consumers of their API(s) with a developer portal. A developer portal provides the platform for an API provider to communicate with the developer community. Addtionally, it typically offers functionality such as user registration and login, user management, documentation, API key management, test console and dashboards. &
$\bullet$ The organization has implemented a developer portal.
& \citedata{de2017api, fremantle2015web, medjaoui2018continuous, sine2015api}    &
  &
6& 
5.4.7 &
Organize Events &
Community Engagement &
Community &
The organization is actively involved in organizing or participating in events that are aimed towards engaging and motivating the developer community to incorporate their API(s) into their applications. This may include events such as hackathons, conferences, or workshops.  &
$\bullet$ The organization is actively involved in organizing or participating in developer community events.
& None.    &
  &
6& 
5.4.9 &
Dedicate Evangelist &
Community Engagement &
Community  &
The organization employs a dedicated API evangelist. This individual is responsible for evangelizing the API by gathering consumer feedback, and promoting the organization's API(s) by creating samples, demos, training materials and performing other support activities aimed towards maximizing the developer experience. &
$\bullet$ The organization employs a dedicated API evangelist.
& None.    &
 &
6& 
5.5.1 &
Enable API Discovery &
Portfolio Management &
Community  &
The organization provides potential consumers of their API(s) with a mechanism to obtain information, such as documentation and metadata, about their API(s). This mechanism may take the shape of an external website, hub or repository that consumers can freely browse through. &
$\bullet$ The organization has a mechanism in place with which their API(s) may be discovered.
& \citedata{biehl2015api, hofman2014technical}    &
 &
6& 
5.5.4 &
Provide API Catalog &
Portfolio Management &
Community  &
The organization provides API consumers with an API Catalog. This is a a searchable catalog of APIs. An API catalog is also sometimes referred to as an API registry. API consumers should be able to search the catalog based on various metadata and tags. The catalog should document the API functionality, its interface, start-up documentation, terms and conditions, reference documentation, and so forth.&
$\bullet$ The organization has implemented the 'Enable API Discovery' (5.5.1) practice. \newline
$\bullet$ The organization provides API consumers with a searchable API catalog.
& \citedata{de2017api, lourencco2019framework, vijayakumar2018practical, hofman2014technical, medjaoui2018continuous}    &
 &
6& 
5.5.5 &
Bundle APIs &
Portfolio Management &
Community  &
The organization is able to combine two or more APIs into a bundle. This is a collection of API products that is presented to developers as a group, and typically associated with one or more rate plans for monetization. &
$\bullet$ The organization is able to combine two or more APIs into a bundle.
&  \citedata{apigeebundling}   &
  &
6&
6.1.1 &
Publish Informal SLA &
Service-Level Agreements
&
Commercial &
The organization has the ability to publish and agree upon an informal, bare-bones SLA with consumers of their API(s). This type of SLA is minimalistic and loose in terms of the nature and amount of agreements it contains, as well as the consequences attached to these agreements should they be violated. This type of SLA is satisfactory for organizations that provide non-critical services and that have close relationships with their consumers and partners. &
$\bullet$ The organization has the ability to publish and agree upon an informal SLA with consumers.
&  None.      &
   &
6& 
6.1.3 &
Provide SLA &
Service-Level Agreements
&
Commercial &
The organization has the ability to provide and agree upon a formal, elaborate SLA with consumers of their API(s). This type of SLA is extensive and strict in terms of the nature and amount of agreements it contains, as well as the consequences attached to these agreements should they be violated. Typically, agreements regarding the guaranteed uptime of the API on a monthly or yearly basis are included in this type of SLA, along with guaranteed response times in the event of incidents, as well as policies regarding privacy, security, and possibly rate and data quotas. Additionally, when providing a formal SLA, the organization should have a plan in place that details what course of action should be taken in the event where agreements are failed to be upheld.
&
$\bullet$ The organization has the ability to provide and agree upon a formal SLA with consumers.
&  \citedata{de2017api}       &
  &
6& 
6.1.6 &
Proactively Monitor SLAs &
Service-Level Agreements
&
Commercial &
The organization is able to proactively monitor metrics that are relevant in checking whether the agreements made with API consumers are adhered to. Such metrics may include availability, performance and functional correctness. &
$\bullet$ The organization has implemented the 'Monitor API Resource Usage' (4.1.5) practice.\newline
$\bullet$ The organization is able to perform SLA monitoring.
& \citedata{moizSLA} &
 &
6& 
6.1.7 &
Customize Personalized SLA &
Service-Level Agreements
&
Commercial &
The organization has the ability to provide consumers of their API(s) with personalized SLAs. This type of SLA is suitable for intensive consumers that utilize services offered by the API in such a way that requires customized agreements as compared to those that are offered as part of the organization's standard SLA. For example, some consumers may require minimal latency and response times for their calls, want to make large amounts of calls, or demand API uptime approaching 100\%. Additionally, a personalized SLA may be required due to the consumer being located in a different geographic location than other consumers, requiring customized agreements with regards to privacy laws and regulations. &
$\bullet$ The organization has implemented the 'Provide SLA' (6.1.3) practice.\newline
$\bullet$ The organization has the ability to provide consumers of their API(s) with personalized SLAs.
& \citedata{manualSLA} &
   &
6& 
6.2.6 &
Adopt Subscription-based Monetization Model &
Monetization Strategy &
Commercial &
The organization has adopted a monetization model that is based on a subscription basis. With this model, API consumers pay a flat monthly fee and are allowed to make a certain number of API calls per month. &
$\bullet$ The organization has implemented the 'Implement Subscription Management System' (6.3.2) and 'Manage Quota' (3.2.5) practices. \newline
$\bullet$ The organization has adopted a monetization model that is based on a subscription basis.
& \citedata{budzynskiMonetization} &
  &
6& 
6.2.8 &
Adopt Tier-Based Monetization Model  &
Monetization Strategy &
Commercial &
The organization has adopted a monetization model that is based on tiered access. Typically, each tier has its own set of services and allowances for access to API resources, with increasing prices for higher tiers. &
$\bullet$ The organization has implemented the 'Prioritize Traffic' (3.2.7) and 'Manage Quota' (3.2.5) practices. \newline
$\bullet$ The organization utilizes a monetization model that is based on tiered access.
&   \citedata{redhatMonetization, budzynskiMonetization}  &
  &
6& 
6.2.9 &
Adopt Freemium Monetization Model &
Monetization Strategy &
Commercial &
The organization has adopted a monetization model that is based on freemium functionalities and access. This involves providing consumers with a limited part of the services and functionalities the API offers as a whole. Consumers that wish to utilize all services and functionalities are required to have an active, paid subscription to the API. 
&
$\bullet$ The organization utilizes a monetization model that is based on freemium functionalities and access.
& \citedata{redhatMonetization, budzynskiMonetization} &
  &
6& 
6.2.10 &
Adopt Metering-Based Monetization Model &
Monetization Strategy &
Commercial &
The organization utilizes a monetization model that is based on metering. With this model, API consumers pay for the amount of resources they use. This may be measured in terms of bandwidth, storage or amount of calls made. &
$\bullet$ The organization has implemented the 'Monitor Resource Usage' (4.1.5) practice.\newline
$\bullet$ The organization utilizes a monetization model that is based on metering.
& \citedata{redhatMonetization, budzynskiMonetization} &
  &
6& 
6.3.2 &
Implement Subscription Management System &
Account Management &
Commercial &
The organization has a system in place with which it is able to manage existing subscriptions (consumers of) on their API. A subscription management system provides support for billing on a recurring basis, as well as providing insight into active subscriptions.
&
$\bullet$ The organization has implemented a subscription management system.
& \citedata{fremantle2015web, preibisch2018api, raivio2011towards}   &
  &
6& 
6.3.7 &
Report on API Program Business Value &
Account Management &
Commercial &
The organization is able to generate business value reports associated with their API(s). Business value reports gauge the monetary value associated with the API program. Monetization reports of API usage provide information on the revenue generated from the API. Value-based reports should also be able to measure customer engagements. Engagements can be measured by the number of unique users, the number of developers registered, the number of active developers, the number of apps built using the APIs, the number of active apps, and many other items. Optionally, these metrics may be visualized in the form of dashboards, so that they may then easily be shared and presented to relevant internal stakeholders to communicate the API program's business value.  &
$\bullet$ The organization has implemented the 'Generate Custom Analysis Reports' (4.3.6) practice. \newline
$\bullet$ The organization is able to generate business value reports associated with their API(s).
& \citedata{de2017api}&
   &
6& 
6.3.8 &
Provide Subscription Report to Customer &
Account Management &
Commercial &
The organization is able to generate subscription reports for consumers of their API(s). These reports contain metrics gathered through internal monitoring and analytics. Such metrics may include amount of calls made, performance, and status regarding remaining allowed quotas. &
$\bullet$ The organization has implemented the 'Generate Custom Analysis Reports' (4.3.6) and 'Implement Subscription Management System' (6.3.2) practices. \newline
$\bullet$ The organization is able to generate subscription reports for consumers of their API(s).
& \citedata{de2017api}&
   &
6& 
6.3.9 &
Proactively Suggest Optimizations to Customers &
Account Management &
Commercial &
The organization has the ability to train and help customers in using their API(s) as well and efficiently as possible. This may be in the best interest of both parties, as optimizing inefficient calls may positively impact traffic load on the API infrastructure. &
$\bullet$ The organization has implemented the 'Monitor API Performance' (4.1.3) and 'Monitor Resource Usage' (4.1.5) practices. \newline
$\bullet$ The organization is able to generate business value reports.
& \citedata{buidesign, de2017api}&
 &
6& 
}
\dataheight=9
\def\returnData(#1){\expandafter\checkMyData(#1)\cachedata}

\newcounter{deTeller}
\newcounter{volgendeStart}
\newcounter{volgendeStop}

\setcounter{deTeller}{1}
\setcounter{volgendeStart}{\value{deTeller}}

\newcounter{tempCount}
\newcounter{groteLoop}
\newcounter{loop}
\newcounter{loopPlusEen}
\newcounter{loopMinEen}

\newcounter{stopTeller}
\newcounter{oldStopTeller}
\newcommand{\lastColumnsWidth}{15.5cm}

\forloop{groteLoop}{1}{\value{groteLoop}<21}{
    \setcounter{oldStopTeller}{0}
    \setcounter{stopTeller}{4} 

    \ifnum\value{deTeller} > \value{oldStopTeller}   \setcounter{volgendeStop}{\value{stopTeller}} \fi
    \setcounter{oldStopTeller}{\value{stopTeller}}
    \addtocounter{stopTeller}{3} 
    \ifnum\value{deTeller} > \value{oldStopTeller}   \setcounter{volgendeStop}{\value{stopTeller}} \fi
    \setcounter{oldStopTeller}{\value{stopTeller}}
    \addtocounter{stopTeller}{4} 
    \ifnum\value{deTeller} > \value{oldStopTeller}   \setcounter{volgendeStop}{\value{stopTeller}} \fi
    \setcounter{oldStopTeller}{\value{stopTeller}}
    \addtocounter{stopTeller}{3} 
    \ifnum\value{deTeller} > \value{oldStopTeller}   \setcounter{volgendeStop}{\value{stopTeller}} \fi
    \setcounter{oldStopTeller}{\value{stopTeller}}
    \addtocounter{stopTeller}{4} 
    \ifnum\value{deTeller} > \value{oldStopTeller}   \setcounter{volgendeStop}{\value{stopTeller}} \fi
    \setcounter{oldStopTeller}{\value{stopTeller}}
    \addtocounter{stopTeller}{6} 
    \ifnum\value{deTeller} > \value{oldStopTeller}   \setcounter{volgendeStop}{\value{stopTeller}} \fi
    \setcounter{oldStopTeller}{\value{stopTeller}}
    \addtocounter{stopTeller}{2} 
    \ifnum\value{deTeller} > \value{oldStopTeller}   \setcounter{volgendeStop}{\value{stopTeller}} \fi
    \setcounter{oldStopTeller}{\value{stopTeller}}
    \addtocounter{stopTeller}{4} 
    \ifnum\value{deTeller} > \value{oldStopTeller}   \setcounter{volgendeStop}{\value{stopTeller}} \fi
    \setcounter{oldStopTeller}{\value{stopTeller}}
    \addtocounter{stopTeller}{7} 
    \ifnum\value{deTeller} > \value{oldStopTeller}   \setcounter{volgendeStop}{\value{stopTeller}} \fi
    \setcounter{oldStopTeller}{\value{stopTeller}}
    \addtocounter{stopTeller}{3} 
    \ifnum\value{deTeller} > \value{oldStopTeller}   \setcounter{volgendeStop}{\value{stopTeller}} \fi
    \setcounter{oldStopTeller}{\value{stopTeller}}
    \addtocounter{stopTeller}{4} 
    \ifnum\value{deTeller} > \value{oldStopTeller}   \setcounter{volgendeStop}{\value{stopTeller}} \fi
    \setcounter{oldStopTeller}{\value{stopTeller}}
    \addtocounter{stopTeller}{5} 
    \ifnum\value{deTeller} > \value{oldStopTeller}   \setcounter{volgendeStop}{\value{stopTeller}} \fi
    \setcounter{oldStopTeller}{\value{stopTeller}}
    \addtocounter{stopTeller}{4} 
    \ifnum\value{deTeller} > \value{oldStopTeller}   \setcounter{volgendeStop}{\value{stopTeller}} \fi
    \setcounter{oldStopTeller}{\value{stopTeller}}
    \addtocounter{stopTeller}{3} 
    \ifnum\value{deTeller} > \value{oldStopTeller}   \setcounter{volgendeStop}{\value{stopTeller}} \fi
    \setcounter{oldStopTeller}{\value{stopTeller}}
    \addtocounter{stopTeller}{3} 
    \ifnum\value{deTeller} > \value{oldStopTeller}   \setcounter{volgendeStop}{\value{stopTeller}} \fi
    \setcounter{oldStopTeller}{\value{stopTeller}}
    \addtocounter{stopTeller}{5} 
    \ifnum\value{deTeller} > \value{oldStopTeller}   \setcounter{volgendeStop}{\value{stopTeller}} \fi
    \setcounter{oldStopTeller}{\value{stopTeller}}
    \addtocounter{stopTeller}{3} 
    \ifnum\value{deTeller} > \value{oldStopTeller}   \setcounter{volgendeStop}{\value{stopTeller}} \fi
    \setcounter{oldStopTeller}{\value{stopTeller}}
    \addtocounter{stopTeller}{4} 
    \ifnum\value{deTeller} > \value{oldStopTeller}   \setcounter{volgendeStop}{\value{stopTeller}} \fi
    \setcounter{oldStopTeller}{\value{stopTeller}}
    \addtocounter{stopTeller}{4} 
    \ifnum\value{deTeller} > \value{oldStopTeller}   \setcounter{volgendeStop}{\value{stopTeller}} \fi
    \setcounter{oldStopTeller}{\value{stopTeller}}
    \addtocounter{stopTeller}{4} 
    \ifnum\value{deTeller} > \value{oldStopTeller}    \setcounter{volgendeStop}{\value{stopTeller}} \fi

    \setcounter{loopPlusEen}{\value{loop}}
    \setcounter{loopMinEen}{\value{loop}}
    \addtocounter{loopPlusEen}{1}
    \addtocounter{loopPlusEen}{-1}

    \begin{table}[ht!]
    \footnotesize
    \begin{tabular}{|p{.1cm}|p{.1cm}|ll|ll|}
    \hline
    \multirow{15}{*}{\rotatebox[origin=c]{90}{\returnData(\value{deTeller},4)}} &
    \multirow{15}{*}{\rotatebox[origin=c]{90}{\returnData(\value{deTeller},3)}} & 
    \forloop{loop}{\value{volgendeStart}}{\value{loop}<\value{volgendeStop}}{
        \textbf{Practice Code}: & \returnData(\value{deTeller},1) & \textbf{Practice Name}: & \returnData(\value{deTeller},2) \\\cline{3-6}
        &&\multicolumn{4}{p{\lastColumnsWidth}|}{\textbf{\textit{Description: }}\returnData(\value{deTeller},5)}\\\cline{3-6}
    &&\multicolumn{4}{p{\lastColumnsWidth}|}{\textbf{\textit{Implemented when:}} \newline \returnData(\value{deTeller},6)}\\\cline{3-6}
    &&\multicolumn{4}{p{\lastColumnsWidth}|}{Literature: \returnData(\value{deTeller},7)}\\\cline{3-6}
    &&\multicolumn{4}{|p{\lastColumnsWidth}}{}\\\cline{3-6}
        &&
        \addtocounter{deTeller}{1}
    } 
    \setcounter{volgendeStart}{\value{deTeller}}

    \textbf{Practice Code}: & \returnData(\value{deTeller},1) & \textbf{Practice Name}: & \returnData(\value{deTeller},2) \\\cline{3-6}
    &&\multicolumn{4}{p{\lastColumnsWidth}|}{\textbf{\textit{Description: }}\returnData(\value{deTeller},5)}\\\cline{3-6}
    &&\multicolumn{4}{p{\lastColumnsWidth}|}{\textbf{\textit{Implemented when:}} \newline \returnData(\value{deTeller},6)}\\\cline{3-6}
    &&\multicolumn{4}{p{\lastColumnsWidth}|}{Literature: \returnData(\value{deTeller},7)}\\\hline
    \end{tabular}
    \end{table}
    \addtocounter{deTeller}{1}
 
} 


\newpage

\section{Version 0.1}
\label{sec:version01}

This version was populated using the primary source~\cite{de2017api}.
It consisted of four focus areas.
Further details are omitted because of the intermediate state of the model.

\begin{table}[h]
    \centering
    \begin{tabular}{l|c}
        Focus Area & Number of capabilities\\
        \hline
        \textbf{Developer Enablement} & 4 \\
        \textbf{Security and Communication} & 5 \\
        \textbf{Lifecycle} & 2 \\
        \textbf{Auditing and Analysis} & 3 \\
    \end{tabular}
    \caption{API-m-FAMM version 0.1}
    \label{tab:version01}
\end{table}

\section{Version 0.2}
\label{sec:version02}

This version was populated using the SLR~\cite{mathijssen2020identification}.
The re-location of practices and capabilities was primarily driven by the decision to split the \textit{security and communication} focus area up into two separate focus areas: \textit{security} and \textit{communication}. 
This decision was made because security was found to be an substantial and integral topic of API management in itself. 
Moreover, it was decided that the communication focus area, which was later renamed to \textit{performance}, comprises capabilities such as \textit{service routing} that are unrelated to security. 
Furthermore, the decision was made to split the \textit{auditing and analytics} focus area up into technical management, which was later renamed to \textit{monitoring}, and business-side, which was later renamed to \textit{commercial}. 
This was done due to the difference in nature between capabilities such as \textit{monetization} and \textit{analytics}, which were originally grouped together. 
This difference was further compounded by the decision to split the traffic management capability into two separate capabilities, with one capturing the business-level aspect of this capability and the other encompassing operational aspects. 
The former capability was then moved to the new commercial focus area along with the monetization capability, while the latter was moved to the performance focus area.

\begin{table}[h]
    \centering
    \begin{tabular}{l|c}
        Focus Area & Number of capabilities\\
        \hline
        \textbf{Community Engagement} & 4 \\
        \textbf{Security} & 2 \\
        \textbf{Communication} & 2 \\
        \textbf{Lifecycle} & 5 \\
        \textbf{Technical Management} & 4 \\
        \textbf{Business Side} & 3 \\
    \end{tabular}
    \caption{API-m-FAMM version 0.2}
    \label{tab:version02}
\end{table}

\section{Version 0.3}
\label{sec:version03}

More information was needed to determine whether practices and capabilities were suited to be included in the model with regards to their scope and relevance. 
In order to resolve this, the collection of practices and capabilities was verified by using information gathered from grey literature such as online blog posts, websites, commercial API management platform documentation and third-party tooling. 
Doing so resulted in the following changes made with regards to the contents of the API-m-FAMM:

\begin{itemize}
    \item \textit{Removal} of several practices that were found to be irrelevant, redundant, or too granular. For example, \textit{filtering spam calls}, which was originally uncovered as part of the SLR, was found to be redundant as this practice is already covered by practices such as \textit{DoS protection} and \textit{rate limiting}. Consequently, such practices were removed.
    \item \textit{Addition} of several practices that were newly identified. For example, \textit{predictive analytics} was found to be a practice that is offered by multiple commercial API management platform providers. Similarly, \textit{including change logs} was found to be a practice that is recommended by practitioners as a best practice when updating APIs. Consequently, such practices were added to the API-m-FAMM.
    \item \textit{Merging} of several practices that were found to be irrelevant, redundant, or too granular. For example, practices that were originally uncovered through the SLR, such as \textit{email-based support}, \textit{phone-based support}, and \textit{form-based support} were found to be redundant, as no significant difference with regards to their maturity may be discerned among these practices. Consequently, these practices were merged into one practice: \textit{establish communication channel}.
    \item \textit{Splitting} of practices that were found to be compounded by practices that were thought to warrant separate, individual practices. For example, the \textit{black or whitelist IP addresses} was split up into the \textit{blacklist IP addresses} and \textit{whitelist IP addresses} practices because these were found to be relevant practices on their own. Additionally, Consequently, these practices were merged into one practice: \textit{establish communication channel}.
    \item \textit{Relocation} of practices to different capabilities than those they were originally assigned to. For example, the \textit{Oauth2.0 authorization} practice was moved from the \textit{authentication} capability to the newly introduced \textit{authorization} capability as Oauth is considered to be an authorization protocol.
    \item \textit{Renaming} of several practices, as well as updating descriptions and formulation of practice descriptions that were previously missing or incomplete. For example, the \textit{provide code samples} practice was renamed to \textit{provide FAQ with code samples} because it was found that these two practices often go hand in hand. Additionally, this practice's description was updated.
    \item \textit{Identification} of dependencies among practices, either among practices within the same capabilities or among practices across different capabilities or focus areas. Some dependencies were found to be relatively straightforward, such as the \textit{multiple API versioning strategy} practice depending on the implementation of the \textit{maintain multiple APIs} practice. However, dependencies between practices belonging to different capabilities such as \textit{quota management} depending on \textit{rate limiting} or \textit{rate throttling} were also identified. 
    \item \textit{Arrangement} of practices based on their interrelated maturity with regards to the other practices in the capability they are assigned to. At this point in time, this was performed on a mostly subjective and empirical basis, and thus should be regarded as a first attempt to discern practices with regards to their relative maturity. 
    \item \textit{Formulation} of implementation conditions corresponding to each practice, which are aimed at providing practitioners with an overview of the necessary conditions that must be met before a practice may be marked as implemented.
\end{itemize}

The amount of practices and capabilities that were added, removed, merged, split, relocated or renamed as a result of the supplemental material validation process and the aforementioned discussion session are shown in Table~\ref{tab:ResultsSupplemental} below. 
However, it should be noted that some practices that were added as a result of the online verification process were later removed as a result of the discussion session. 
As such, numbers corresponding to the \textit{added} and \textit{removed} operations presented in Table~\ref{tab:ResultsSupplemental} are slightly inflated.

\begin{table}[h]
\centering
\begin{tabular}{l|c|c|c|c|c|c}
\textbf{Component} & \textbf{Added} & \textbf{Removed} & \textbf{Merged} & \textbf{Split} & \textbf{Relocated} & \textbf{Renamed}\\
\hline
Practice & 17 & 27 & 39 & 4 & 12 & 93 \\ 
Capability & 1 & 1 & 1 & 0 & 1 & 2 \\
\end{tabular}
\caption{Number of practices and capabilities added, removed, merged, split, relocated or renamed as a result of the supplemental material validation process and the discussion session.}
\label{tab:ResultsSupplemental}
\end{table}

\begin{figure*} 
\centering
\includegraphics[page=1, clip, trim=0cm 0cm 0cm 0cm, width=\textwidth]{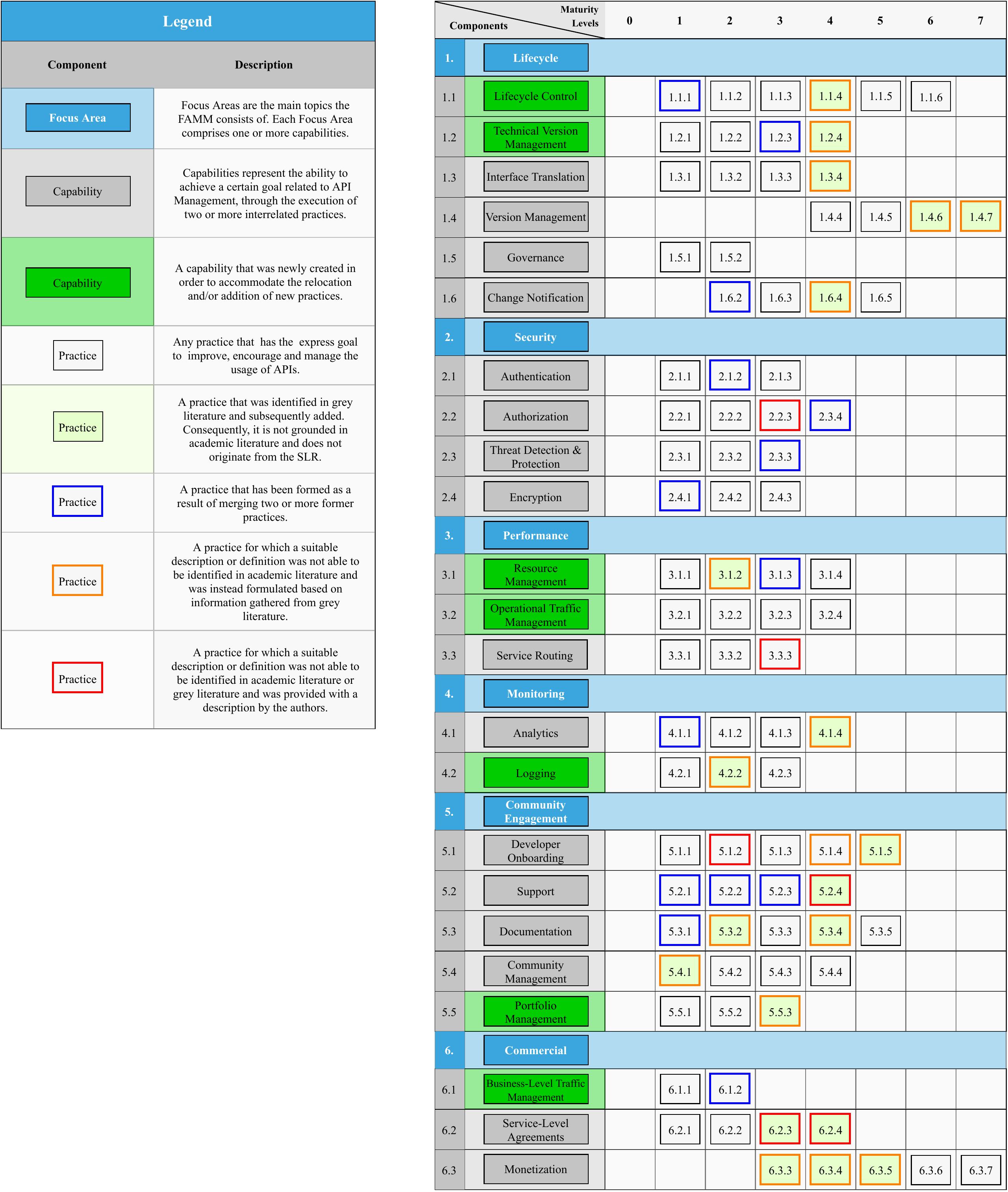}
\caption{version 0.3 of the API-m-FAMM and the focus areas, capabilities, and practices it consists of. Additionally, it is shown which capabilities and practices were newly introduced between API-m-FAMM v0.2 and v0.3, as well as for which practices descriptions were formulated based on supplemental material. Please consult the legend on the top left-hand side of the figure for more information regarding the differently shaped and/or colored components.}
\label{fig:api-m-famm03}
\end{figure*}

At this stage of the design process, the model is grounded in literature, and is verified and supplemented by using grey literature. 
As a result of these activities, the initial body of 114 practices and 39 capabilities that was extracted as a result of the SLR was refined and narrowed down to 87 practices and 23 capabilities, which are divided among six focus areas. 
Instead, the contents of this version of the API-m-FAMM can be found in \emph{version2} of this published source document on arXiv~\cite{mathijssen2021source}. 
The general structure of the API-m-FAMM version 0.3 is presented in Figure~\ref{fig:api-m-famm03}. As shown, each individual practice is assigned to a maturity level within its respective capability. Additionally, it should be noted that practices can not depend on practices as part of another capability that have a higher maturity level. For example, practice 1.4.4 is dependant on the implementation of practice 1.2.3, resulting in a higher maturity level being assigned to the former of these practices.

Figure~\ref{fig:api-m-famm03} also shows that at this stage, 17 practices were added in addition to those extracted through the SLR. Furthermore, 14 new practices were introduced as a result of merging 39 former practices, as shown in Table~\ref{tab:ResultsSupplemental}. Moreover, descriptions that are based on grey literature were formulated for 18 practices for which adequate descriptions were not able to be identified in academic literature. Lastly, 6 practices are accompanied by descriptions that were formulated by the researchers themselves, as based on empirical knowledge. Even though suitable descriptions could not be identified for these practices in academic literature or grey literature, they were included in this version of the API-m-FAMM because they were hypothesized to be relevant for practitioners. Among other things, this hypothesis is tested through expert interviews, which are part of the next phase in constructing the API-m-FAMM.

\section{Version 0.4}
\label{sec:version04}

\begin{figure*} 
\centering
\includegraphics[page=1, clip, trim=7cm 4cm 9cm 0.5cm, width=0.8\textwidth]{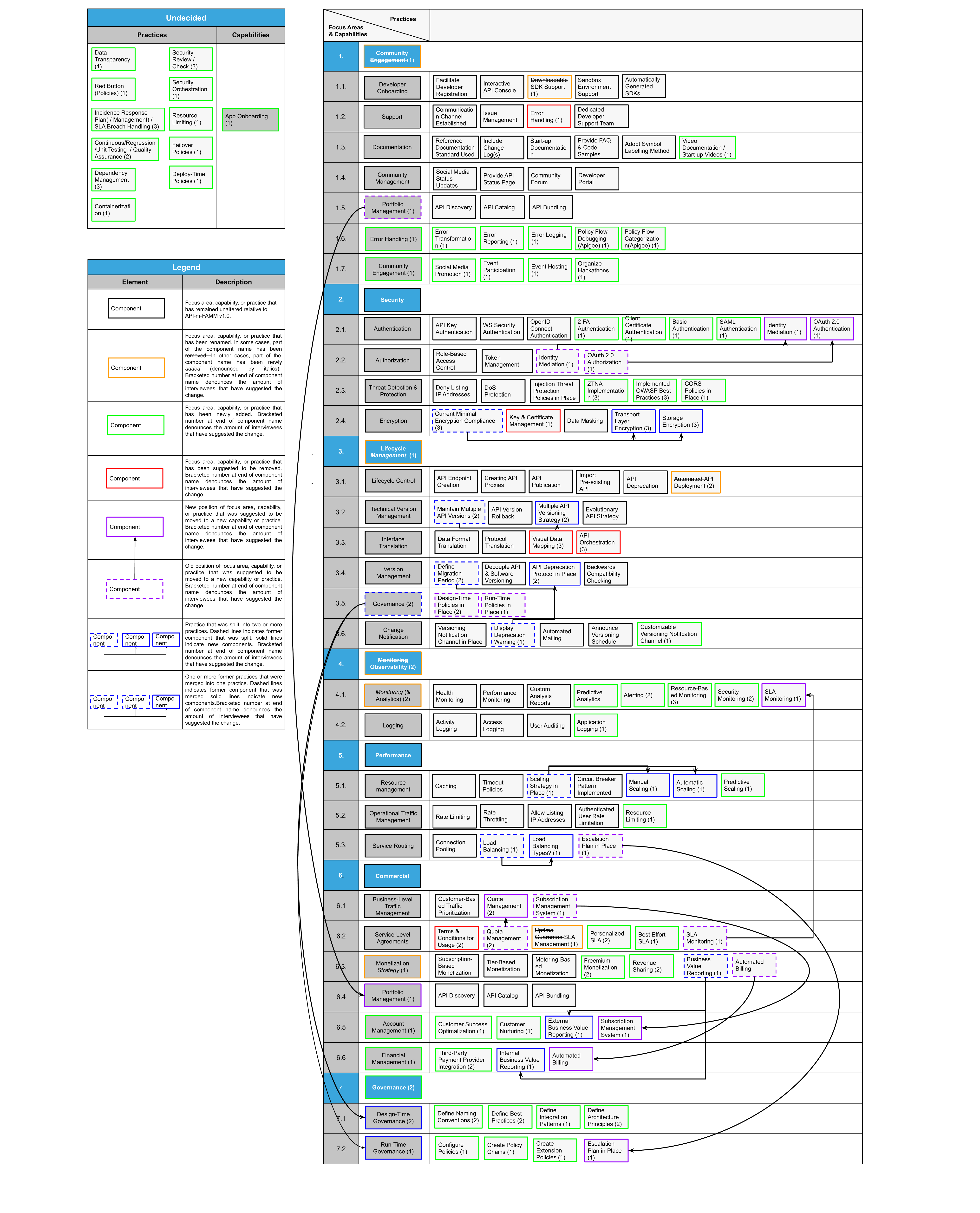}
\caption{API-m-FAMM version 0.3 plus all suggested changes that were made by experts during interviews. Please consult the legend on the left-hand side of the figure for more information regarding the manner in which the colored outlines should be interpreted. Practices and capabilities that were not directly categorized by the expert during interviews are placed in the 'undecided' box on the top-left hand side.}
\label{fig:api-m-famm04a}
\end{figure*}

Eleven expert interviews were conducted. 
During these interviews, many additions and changes in terms of the API-m-FAMM's structure and contents were suggested by experts, whom were encouraged to elaborate on their motivation regarding these suggestions. 
By transcribing and processing the recordings of all interviews, the numerous suggestions that were made by experts to either add, remove, merge, split, relocate, or rename several focus areas, capabilities, and practices, are compiled. 
The amount in which these suggestions for changes occurred are shown in Table \ref{tab:EvaluationChanges} below, as grouped by the type of suggested change as well as the type of component they apply to. Additionally, these changes are visually represented in their entirety in Figure \ref{fig:api-m-famm04a}, along with the number of experts that suggested for a specific change to be made. Evidently, the number of practices that were suggested to be added is relatively high. It should be noted that while a large part of these practices were explicitly mentioned by experts, some were also indirectly extracted from transcripts as a result of comments the expert had made. Additionally, no suggestions are rejected at this point, hence all suggestions that were made by experts are taken into account and incorporated into Table \ref{tab:EvaluationChanges} and Figure \ref{fig:api-m-famm04a}.

\begin{table}[h]
\centering
\begin{tabular}{l|c|c|c|c|c|c}
\textbf{Component} & \textbf{Added} & \textbf{Removed} & \textbf{Merged} & \textbf{Split} & \textbf{Relocated} & \textbf{Renamed}\\
\hline
\textbf{Practice} & 50 & 5 & 3 & 3 & 9 & 3 \\
\textbf{Capability} & 7 & 0 & 0 & 2 & 2 & 2 \\
\textbf{Focus Area} & 1 & 0 & 0 & 0 & 0 & 3\\
\end{tabular}
\caption{Number of practices, capabilities, and focus areas that were suggested to be added, removed, merged, split, relocated or renamed by experts during interviews.}
\label{tab:EvaluationChanges}
\end{table}

\begin{figure*}[!h]
\centering
\includegraphics[page=1, clip, trim=7cm 4cm 2cm 0.5cm, width=0.8\textwidth]{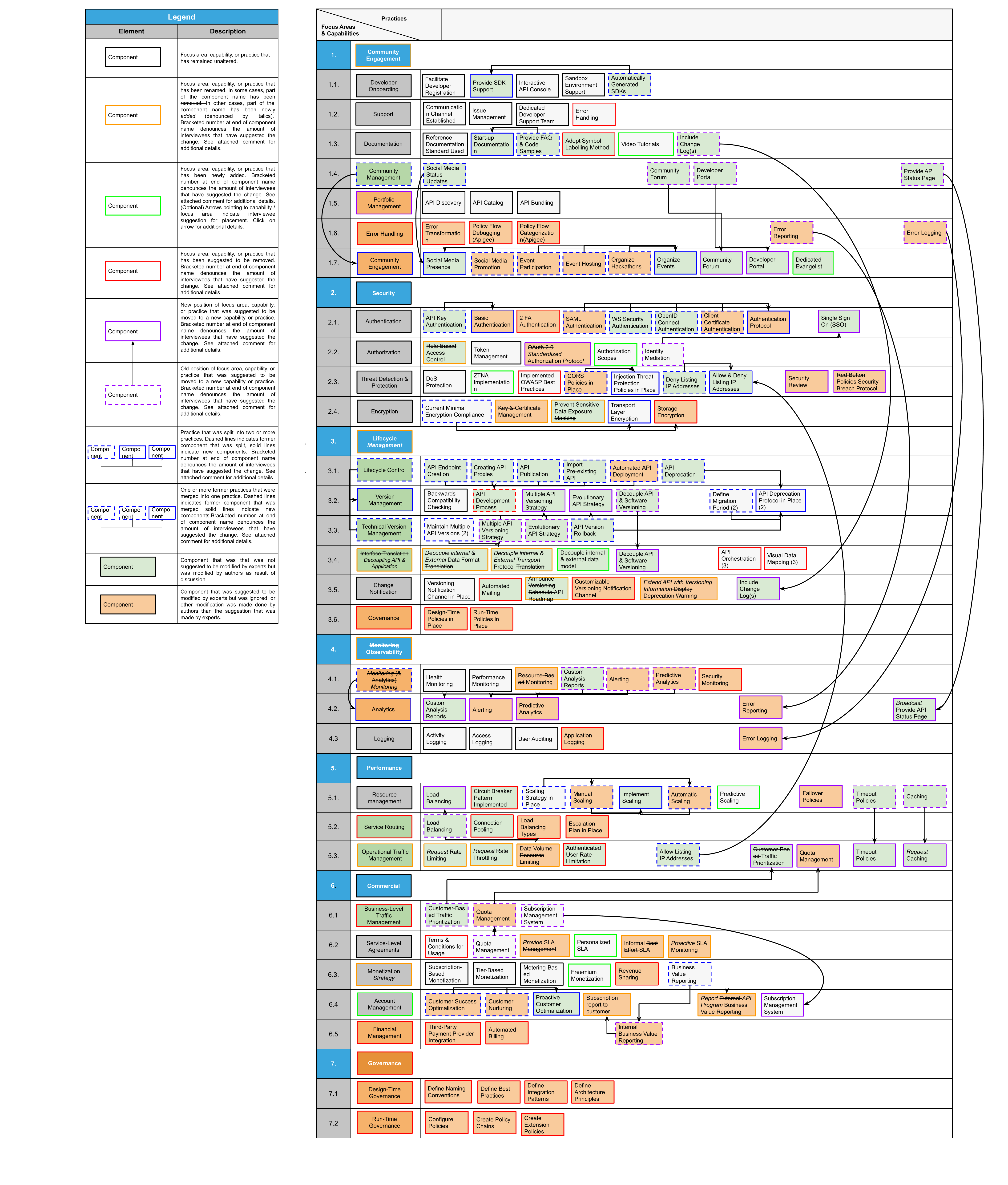}
\caption{API-m-FAMM v0.4, including all suggested changes that were made by experts during interviews, as well as the manner in which they were subsequently interpreted and applied by the researchers. Please consult the legend on the top left-hand side of the figure for more information regarding the manner in which the colored outlines and fills should be interpreted.}
\label{fig:api-m-famm04b}
\end{figure*}

After having compiled all suggestions made by experts, extensive discussion sessions are held among all authors to analyze, discuss, and interpret them. 
All suggested changes to either a focus area itself, or the capabilities or practices it consists of are then analyzed and interpreted through the help of the transcribed arguments that were provided by experts during the interviews. 
As a result, numerous modifications are made to the API-m-FAMM, which are visualized in its entirety in Figure \ref{fig:api-m-famm04b}. 
Additionally, some fundamental decisions are made with regards to the scope and contents of the API-m-FAMM.

\begin{itemize}
    \item Firstly, it was decided that all practices that are contained in the model should be implementable \textit{without} the usage of an API management platform. This decision was made due to several reasons. First of all, it was found that among the organizations that the experts that were consulted are employed at, only a small portion actively utilizes a third party platform to manage their API(s). When asked, experts belonging to the category that have not incorporated an API management platform into their organizations cited arguments such as wanting to avoid vendor lock-in, high costs, or simply not having a need for many of the functionalities provided by such management platforms. Oftentimes, the latter argument was tied to the organization currently exclusively using internal APIs, thus removing the need for using a management platform to manage and expose any partner or public APIs altogether. Considering that it may reasonably be hypothesized that these arguments may likely also apply to other organizations wishing to consult the API-m-FAMM to evaluate and improve upon their API management related practices, any practices or capabilities that were found to be directly tied to the usage of an API management platform were removed from the model. For example, this was the case for the \textit{Visual Data Mapping} practice, which is exclusively provided by the \textit{Axway} API management platform \footnote{\url{https://www.axway.com/en/products/api-management}}, as well as the practices corresponding to the newly suggested \textit{Error Handling} capability, which are implementable through the use of the \textit{Apigee} platform \footnote{\url{https://cloud.google.com/apigee/api-management?hl=nl}}.
    
    An additional reason for excluding such capabilities and practices is that they are likely to evolve throughout the coming years, which would in turn require the API-m-FAMM to be updated as well. In order to prevent this, the API-m-FAMM and the practices it comprises should be platform-independent. Lastly, the purpose of the API-m-FAMM is not to guide practitioners in selecting an appropriate commercial API management platform for their organization. Instead, the API-m-FAMM aims to guide organizations in assessing and evaluating their current maturity in terms of those processes that are considered to be best-practices and are at the core of API management, so that they may then develop a strategy towards implementing practices that are currently not implemented and desirable in further maturing the organization in terms of API management.

    \item Secondly, many practices were deemed to be too granular, specific, or irrelevant to be included. Consequently, such practices were either removed, or merged into a practice that is composed of these smaller practices. An example of practices that were found to be too granular include newly suggested practices such as \textit{Event Participation}, \textit{Event Hosting}, and \textit{Organize Hackathons}. Additionally, since determining a difference among these practices in terms of their maturity was found to be unfeasible, they were instead merged into the \textit{Organize Events} practice and included in its description.

    \item Thirdly, some practices that describe a specific protocol were renamed to be more ambiguous and generic. For example, the former \textit{OAuth 2.0 Authorization} practice was renamed to \textit{Standardized Authorization Protocol}, with a referral to the OAuth 2.0 protocol being included in its description instead. This was done to ensure that the API-m-FAMM remains functional and applicable in the future, since it is likely that new protocols will be developed and adopted among the industry in the future. These concerns also applied to suggested practices corresponding to individual authentication methods such as client certificate and SAML authentication, which were ultimately merged into the \textit{Implement Authentication Protocol} practice and included in its description. An additional reason for doing so in the case of these authentication methods is that they each have their individual strengths and weaknesses, with one not always necessarily being 'better' or more mature than the other. Furthermore, some methods may be more appropriate for some use cases than others. 
    
    \item Furthermore, some capabilities and its corresponding practices that were also thought to apply to most organizations in general, that are not necessarily involved with API management were excluded from the model. An example of this is the \textit{Financial Management} capability that was suggested to be added. Considering that practices such as \textit{Automated Billing}, \textit{Third-Party Payment Provider Integration}, and \textit{Revenue Sharing} are best practices that apply to commercially oriented organizations in general, they were removed. This decision was made to ensure that the contents of the API-m-FAMM is exclusively composed of practices that are directly tied to API management.

    \item During interviews focused on the \textit{Lifecycle} focus area, experts were asked to elaborate on the manner in which their organization has implemented \textit{Governance}. Based on the answers given however, it became clear that capturing processes related to governance in the form of practices is not feasible. This may largely be attributed to the observation that such processes seem to be inherent to specific characteristics of the organization, such as its culture, size, usage of a third party API management platform, as well as the amount of APIs that are used or exposed by the organization.
    
    Some practices were suggested for addition, such as \textit{Define Naming Conventions}, \textit{Define Best Practices}, and \textit{Define Integration Patterns}. However, after having discussed these with experts in subsequent interviews, it was decided that these practices are too abstract and inconcrete in comparison with other practices, considering that they may be interpreted in different ways by practitioners due to the varying organizational characteristics mentioned earlier. Hence, the \textit{Governance} capability that was originally part of the \textit{Lifecycle} focus area was removed, along with the \textit{Design-time Governance} and \textit{Run-time Governance} practices it was composed of.

    \item A valuable suggestion that was made by experts is the addition of monitoring in terms of the amount of resources that calls to the API consume, such as CPU, disk, memory, and network usage. Considering that this monitoring perspective was previously missing alongside performance and health monitoring, as well as it being suggested by multiple experts independently from one another, the \textit{Resource Monitoring} practice was newly added. Similarly, this resource perspective was also found to be missing among the \textit{Traffic Management} capability, alongside the \textit{Request Limiting} and \textit{Request Throttling} practices. Hence, the \textit{Data Volume Limiting} practice was newly added.

    \item Another fundamental change that was made to the API-m-FAMM is the renaming of the former \textit{Monitoring} focus area to \textit{Observability}. This rename was independently suggested by two experts, whom argued that observability better describes the focus area, considering that the \textit{Analytics} capability was split into two capabilities: \textit{Monitoring} and \textit{Analytics}. This decision was made because experts were of the opinion that monitoring is concerned with gathering (real-time) metrics related to the API's health, performance, and resource usage, while analytics is concerned with aggregating these metrics so that insights may be formed and subsequent action may be taken based off of these. As a result, the monitoring capability was added, as well as practices related either to monitoring or analytics being moved to the capabilities they are associated with. 
    
    \item  Moreover, some practices that were originally posed from a passive perspective, were changed with the intention of being conducted in an active manner. For example, the \textit{Include Changelogs} practice was renamed to \textit{Distribute Changelogs}, and its description was changed so that its focus is changed from passive inclusion of changelogs in the reference documentation, to active distribution of changelogs to consumers of the API. Similarly, the \textit{Provide API Status Page} was renamed to \textit{Broadcast API Status}, as well as its description being changed to signify the operational status of the API being broadcasted to consumers in an active manner, as opposed to providing an API status page in a passive fashion. These changes were made due to the fact that when phrased in a passive manner, these practices were deemed to be too irrelevant to be included in the API-m-FAMM, considering that the level of maturity required to implement these practices is too low when compared to other practices. When phrased from an active perspective however, these practices can be considered to be best practices that an organization should strive to implement.

    \item Finally, a major fundamental change was made with regards to the \textit{Lifecycle Control} capability. While practices belonging to this capability such as \textit{API Endpoint Creation}, \textit{API Publication}, and \textit{Import Pre-existing API} are considered to be an integral aspect of API management in both literature as well as the industry, the decision was made to exclude these practices from the API-m-FAMM. This choice was made due to the fact that being able to design, create, publish, and deploy an API is a precondition for implementing all other practices the model consists of. Moreover, during interviews it became clear that it was difficult for experts to rank these practices in terms of their maturity, considering that they are often performed in chronological order.
    \end {itemize}

\begin{figure*}[!h]
\centering
\includegraphics[page=1, clip, trim=1cm 3cm 8cm 1cm, width=0.7\textwidth]{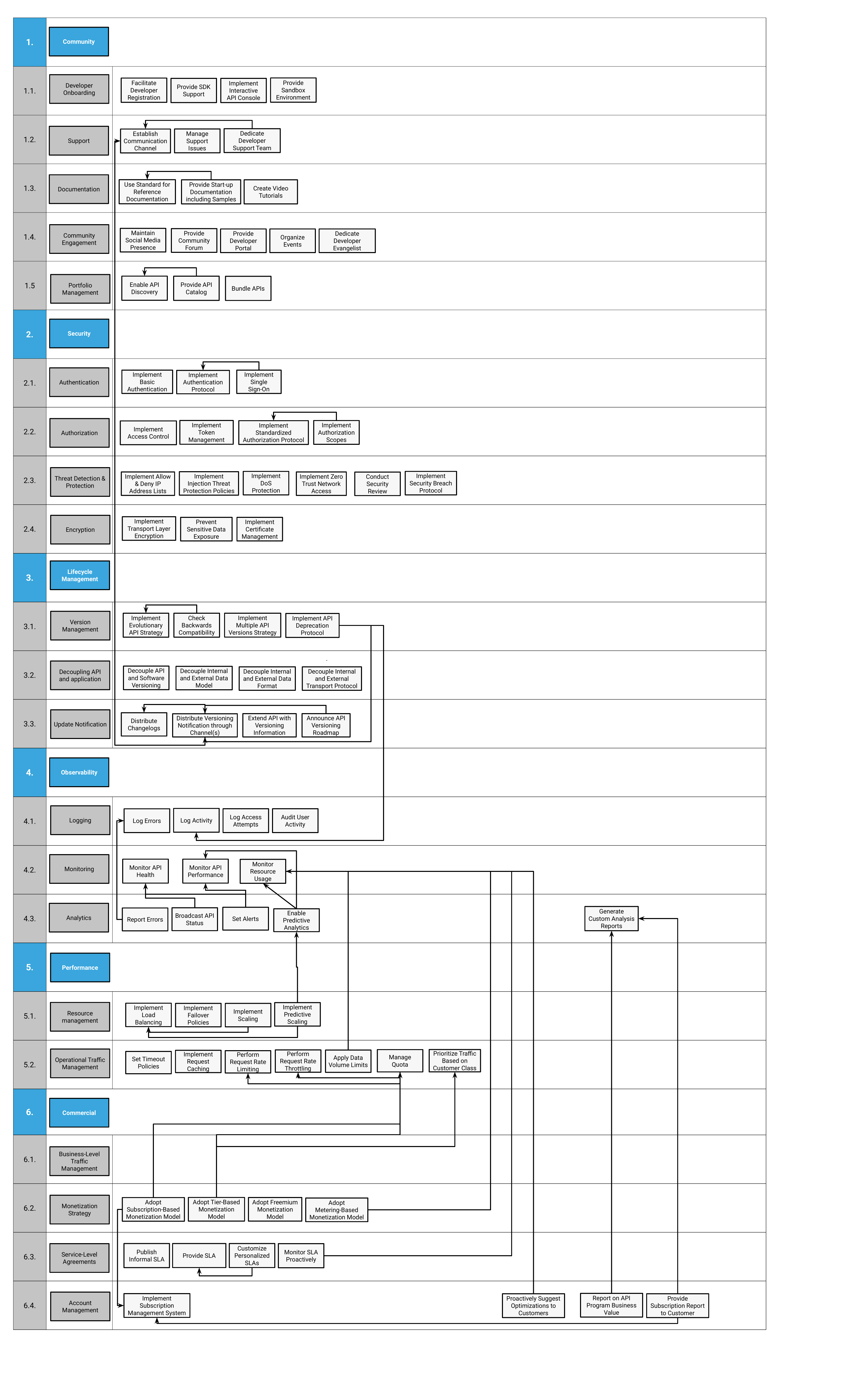}
\caption{The API-m-FAMM v0.4 after all changes had been applied, showing all dependencies that were identified between practices. In order to improve legibility, practices are not ranked in terms of their maturity in this figure.}
\label{API-m-FAMM Dependencies}
\end{figure*}

Next the practices are assigned to individual maturity levels.
This is done by using the results of the maturity ranking exercises during the interviews. 
First however, all dependencies between practices are identified, which are depicted in Figure \ref{API-m-FAMM Dependencies}. 
In this context, a dependency entails that one or more practices that the practice in question is dependant on are required to be implemented before the practice may be implemented. 
These dependencies may either occur; (1) between practices within the same capability; (2) between practices that are assigned to different capabilities within the same focus area, or (3) between practices that are assigned to different capabilities and focus areas. 
In total 34 dependencies are identified, which was done by analyzing literature stemming from the SLR and online supplemental material, as well as input received through expert interviews and the discussion sessions that were conducted among the researchers. The number of dependencies that are identified are shown for each focus area in Table \ref{tab:DependenciesTable}, as well as for each of the three dependency types mentioned.

\begin{table}[h]
\centering
\begin{tabular}{l c c c r}
\hline
\textbf{Focus Area} &  \textbf{Within Capability} & \textbf{Within Focus Area} & \textbf{Between Focus Areas} & \textbf{Total} \\ 
\hline
Community & 3 & 0 & 0 & 3 \\
Security & 2 & 0 & 0 & 2 \\
Lifecycle Management & 3 & 1 & 2 & 6 \\
Observability & 0 & 6 & 0 & 6 \\
Performance & 4 & 0 & 2 & 6 \\
Commercial & 2 & 1 & 8 & 11 \\
\hline
\textbf{Total} & 14 & 8 & 12 & 34 
\end{tabular}
\caption{The number of identified dependencies per focus area and per dependency type.}
\label{tab:DependenciesTable}
\end{table}

As an example of a dependency between practices within the same capability, implementation of the \textit{Implement Load Balancing} practice is required before the \textit{Implement Scaling} practice may be implemented. 
An example of a dependency between practices that are assigned to different capabilities within the same focus area is the dependency between \textit{Enable Predictive Analytics} and \textit{Performance Monitoring}. The former practice belongs to the \textit{Analytics} capability, while the latter practice belongs to the \textit{Monitoring} capability, but both capabilities are contained within the \textit{Observability} focus area. An example of a dependency between practices that are assigned to different capabilities and focus areas may be observed in the case of the dependency between the \textit{Adopt Metering-based Monetization Model} and \textit{Resource Monitoring} practices. The former practice is assigned to the \textit{Monetization Strategies} capability within the \textit{Commercial} focus area, while the latter practice is assigned to the \textit{Monitoring} capability within the \textit{Performance} focus area.

After having identified all dependencies between practices, all 34 practices that have one or more dependencies are juxtaposed in a matrix. 
This is done by adhering to the constraint that practices can not depend on practices that have a higher maturity level. 
As a result, the foundation of the API-m-FAMM is formed, with practices ranging from maturity levels 1 to 10.
Using this structure as a base, all other practices are subsequently assigned to individual maturity levels within their respective capabilities. 
These assignments are performed by using the results of the maturity ranking exercises that were performed by experts as one of the main sources of input.

\begin{figure}[h]
\centering
\includegraphics[page=1, clip, trim=1cm 0cm 1cm 0cm, width=0.7\textwidth]{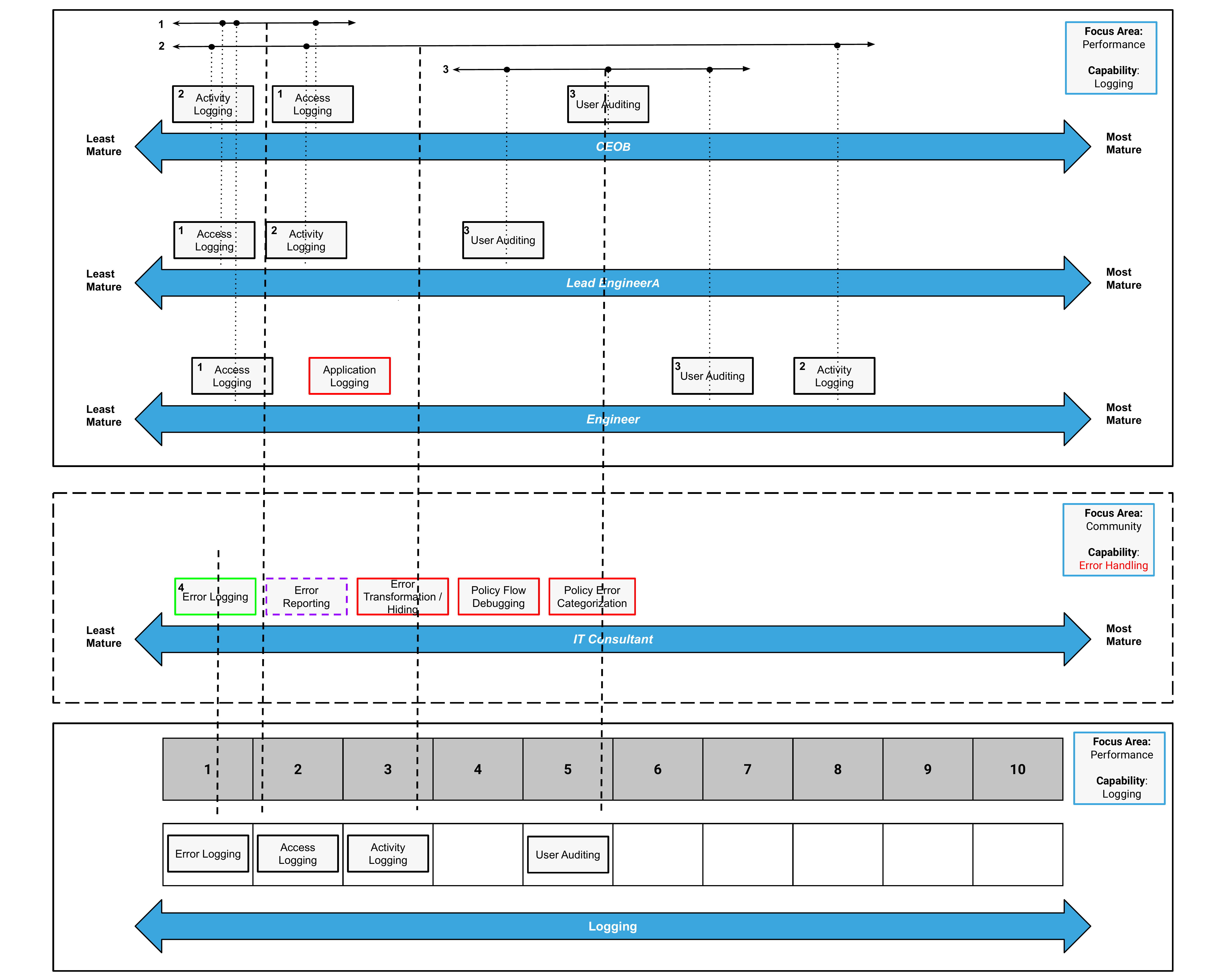}
\caption{Conceptual overview representing a rough approximation of the way in which the expert's maturity rankings were interpreted and used as a starting point for performing the maturity level assignments.}
\label{Maturity_Ranking_Interpretation}
\end{figure}

By again using the \textit{Logging} capability as an example, the interpretation of such a maturity ranking exercise is visualized in Figure \ref{Maturity_Ranking_Interpretation}. 
In this figure, it can be seen that the \textit{Activity Logging}, \textit{Access Logging}, and \textit{User Auditing} practices were ranked by 3 experts in terms of their perceived maturity. 
An additional practice, \textit{Application Logging}, was suggested for addition. 
However, this practice was removed because the decision was made to exclude applications in terms of abstraction from the API-m-FAMM, which is why it is outlined in red. 
Additionally, the decision was made to include and move the \textit{Error Logging} practice to the \textit{Logging} capability. 
Hence, this practice is outlined in green, and is included in this ranking exercise by incorporating this practice in the figure, along with the capability it was originally categorized with by the expert. 
Furthermore, the \textit{Error Reporting} practice was moved to the \textit{Analytics} capability (as can be seen in Figure \ref{fig:api-m-famm04b}, which is why it is outlined in purple and excluded from this maturity ranking exercise. 
Lastly, the remaining 3 practices that were suggested to be added are excluded, along with the \textit{Error Handling} capability as a whole, which is denoted by the red outlines.

Arrows are included that range from the lowest a practice has been ranked in terms of its perceived maturity, to its highest. Dotted lines are attached to each practice, which are then connected to these arrows with a small circle in order to highlight and compare the maturity assignments of each expert with one another. Subsequently, dashed lines are used to indicate a rough estimate of the average of these assignments, which are then mapped on the maturity levels. 
However, it should be noted that Figure \ref{Maturity_Ranking_Interpretation} was made for illustratory purposes, in order to provide the reader with a conceptual idea of the manner in which the maturity assignments were performed. 
In practice, the maturity assignment of practices was done in a pragmatic manner, through discussion sessions among the researchers during which the expert's varying maturity rankings and their accompanying motivation and arguments were discussed and interpreted. Based on the outcome of these discussions, decisions were then made to assign practices to individual maturity levels, while taking the experts' opinions and maturity rankings into account.

\begin{figure*}[!h]
\centering
\includegraphics[page=1, clip, trim=0.5cm 0.5cm 0.5cm 0.5cm, width=\textwidth]{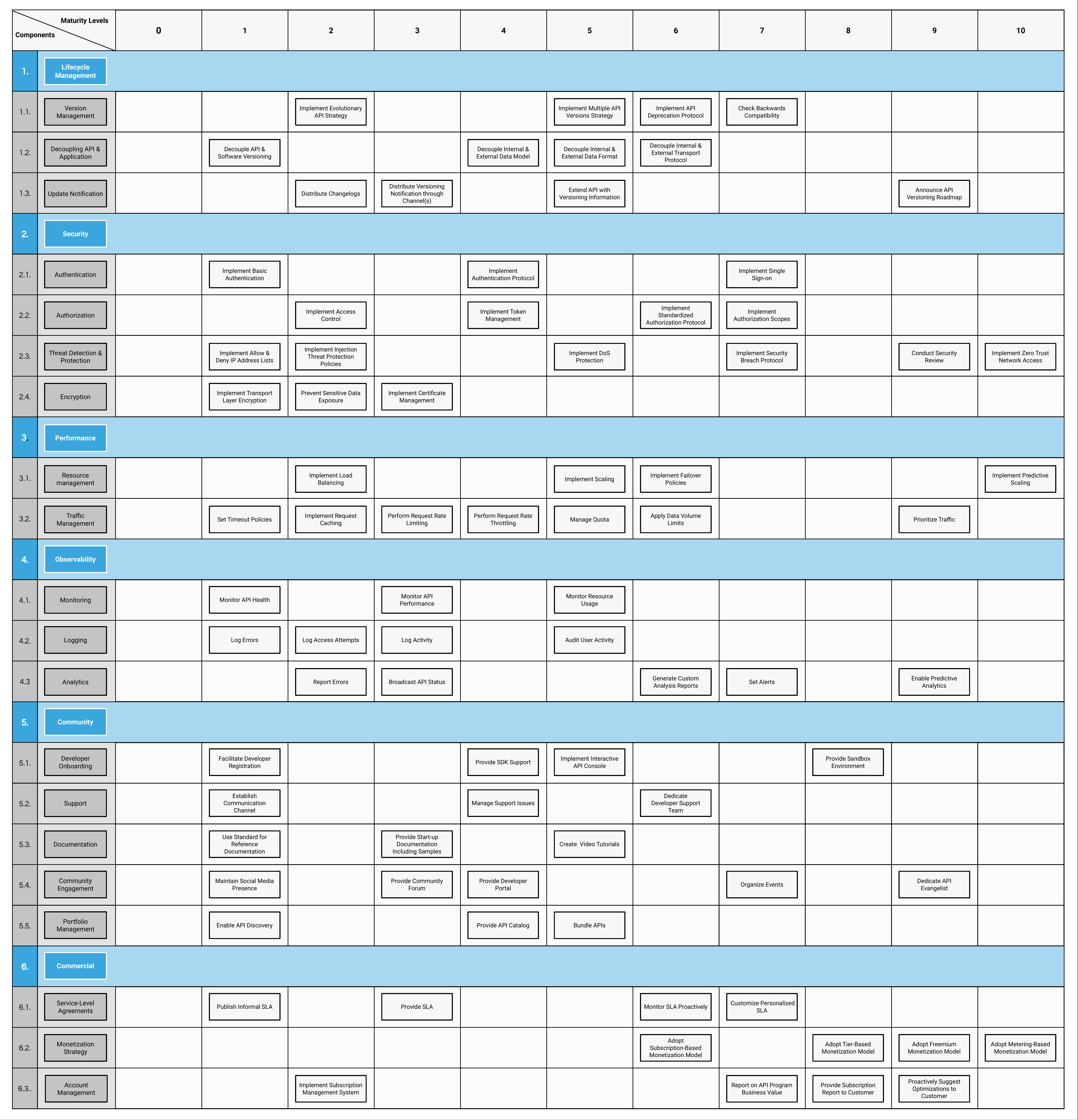}
\caption{API-m-FAMM v0.4, which includes the assignment of all practices to their respective maturity levels, which range from level 1 to level 10.}
\label{API-m-FAMM_2.4}
\end{figure*}

Finally, all practices are renamed to fit an uniform syntactical structure, which starts with a verb, followed by one or more nouns. 
For example, \textit{User Auditing} is renamed to \textit{Audit Users}, and \textit{Resource Monitoring} is renamed to \textit{Monitor Resource Usage}. 
Furthermore, descriptions of the practices that are included in the API-m-FAMM after all changes had been applied are updated. 
When possible, this is done using information and input that was provided by experts during interviews. 
Ultimately, these activities produced a second, updated version of the API-m-FAMM, which is shown in Figure \ref{API-m-FAMM_2.4} and consists of 6 focus areas, 20 capabilities, and 81 practices. 
These descriptions are available through \emph{version3} of this published source document on arXiv~\cite{mathijssen2021source}.

\section{Version 0.5}
\label{sec:version05}

After having updated the API-m-FAMM to incorporate all findings from the interviews a second evaluation cycle was conducted. 
This is done as a means for evaluating and verifying whether experts agree with the fundamental decisions that were made, as well as gathering feedback on the way suggestions made by experts were interpreted and the maturity levels that practices had been assigned to. 
This second evaluation cycle consists of unstructured interviews with three experts originating from the same sample of experts that were interviewed during the first evaluation cycle.
During these interviews, the changes made as a result of the previous evaluation cycle, as well as the newly introduced maturity assignments are presented and discussed.

Since all experts agreed with the fundamental decisions that were made, no major further adjustments are made to the API-m-FAMM as a result of this evaluation cycle.

\section{Version 1.0}
\label{sec:version10}

The final phase of the API-m-FAMM construction, the \emph{Deploy} phase, was executed through case studies.
These case studies were conducted by evaluating six software products.
Some additional changes are made to practices as a result of the discussion sessions with practitioners after the evaluation. 
One practice was removed altogether, and the descriptions of six practices were modified. Specifically, the following changes were made to the following practices:

\begin{itemize}
    \item \textbf{Perform Request Rate Limiting}: this practice was extended to also comprise error limiting. In the case of AFAS Profit, this is implemented by placing consumers on a temporary denylist when they perform an excessive amount of faulty calls within a predefined time span.
    \item \textbf{Prevent Sensitive Data Exposure}: this practice was removed. During discussions, this practice caused confusion due to the observation that this practice is already captured by the \textit{Implement Transport Layer Encryption} and \textit{Decouple Internal \& External Data Model} practices. Additionally, after further investigation this practice was deemed to be out of scope, considering that the scope of this practice involves app data storage in general, as opposed to API management.
    \item \textbf{Implement Predictive Scaling}: the description of this practice was modified. Originally, the description mentioned that this practice may be implemented 'manually or automatically', which caused confusion due to the fact that these methods are already capture in the \textit{Implement Scaling} practice. Because predictive scaling is envisioned by practitioners and the researchers to be done automatically, the manual element was removed from the description.
    \item \textbf{Monitor Resource Usage}: the description of this practice was expanded. During discussions, it became clear that monitoring resources does not always specifically involve metrics such as CPU and disk usage. Instead, rough approximations may be used to determine resource usage instead, which is why the description was expanded to clarify this.
\end{itemize}

In addition to these changes, a small number of changes were made as a result of practitioners identifying errors such as typos. 
The final version of the model can be seen in Figure~\ref{fig:api-m-famm}.

\clearpage

\bibliographystyledata{elsarticle-num}
\bibliographydata{apimanagement}

\clearpage
\bibliographystyle{elsarticle-num}
\bibliography{apimanagement}

\end{document}